\documentclass[twocolumn,amsmath,amssymb,floatfix,phf]{revtex4-2}

\usepackage[english]{babel}
\usepackage[utf8x]{inputenc}
\usepackage{amsmath,amsfonts,amssymb,amsbsy,eucal}
\usepackage{natbib}
\usepackage[usenames, dvipsnames]{color}
\usepackage[colorlinks]{hyperref}
\usepackage{graphicx}
\usepackage{epstopdf}
\usepackage{rotating}
\usepackage{ulem}
\usepackage{bm}  
\usepackage{epsfig}
\usepackage{soul}

\graphicspath{{./}}

\newcommand{\hide}[1]{}

\newcommand{\figurewidth}{0.95\columnwidth}

\begin{document}

\title{Boosting micromachine studies with Stokesian Dynamics }

\author{I. Berdakin}

\affiliation{Facultad de Matemática, Astronomía, Física y Computación, Universidad Nacional de C\'ordoba, C\'ordoba X5000HUA, Argentina}
\affiliation{ Instituto de Física Enrique Gaviola, CONICET - UNC, Córdoba X5000HUA, Argentina}

\author{V. I. Marconi}
\affiliation{Facultad de Matemática, Astronomía, Física y Computación, Universidad Nacional de C\'ordoba, C\'ordoba X5000HUA, Argentina}
\affiliation{ Instituto de Física Enrique Gaviola, CONICET - UNC, Córdoba X5000HUA, Argentina}
           
\author{Adolfo J. Banchio}
              \email{banchio@famaf.unc.edu.ar}   
\affiliation{Facultad de Matemática, Astronomía, Física y Computación, Universidad Nacional de C\'ordoba, C\'ordoba X5000HUA, Argentina}
\affiliation{ Instituto de Física Enrique Gaviola, CONICET - UNC, Córdoba X5000HUA, Argentina}

\begin{abstract}

Artificial microswimmers, nano and microrobots,  are essential in many applications from engineering to biology and medicine.   We present a Stokesian Dynamics  study of the dynamical properties and efficiency of one of the simplest artificial swimmer, {\it the  three linked spheres swimmer} (TLS), extensively shown to be an excellent and model  example of  a deformable micromachine. 
Results for two different swimming strokes are compared with an approximate solution based on point force interactions. While this approximation accurately reproduces the solutions for swimmers with long arms and strokes of small amplitude, it fails when the amplitude of the stroke is such that the spheres come close together, a condition where indeed the largest efficiencies are obtained.  We  find  that swimmers with a ``square stroke cycle'' result more efficient than those with ``circular stroke cycle'' when the swimmer arms are long compared with the sphere radius, but the differences between the two strokes are smaller when the arms of the swimmers are short. This extended theoretical research of TLS incorporates a much precise description of the swimmer hydrodynamics, demonstrating the relevance of considering the finite size of the constitutive microswimmers spheres.  This work expects to trigger future innovative steps  contributing to  the design of  micro and nanomachines and its applications.

\end{abstract}
\maketitle


\section{INTRODUCTION}

Self-propulsion of microorganisms and artificial swimmers is only possible through the generation of motility strategies that are able to overcome the absence of inertia. 
This condition, implied in every low Reynolds number regime, allows the success of only those swimming strategies that are non--reciprocal, {\it i.e.} time--reversed motion is not the same as the original one~\cite{purcell_1977}.  
In the past two decades, there has been a growing interest in understanding the dynamics of self--propelled microorganisms and artificial swimmers.
For recent results and reviews see Refs.~\onlinecite{lauga_2009,elgeti_2015,lauga_2016,Gompper_2020}, and references therein.

Artificial microswimmers, micromachines and nano--robots are of great  present
interest for technical and medical
applications~\cite{nelson_2010,li_2017,hu_2018,ghosh_2020}, like 
cargo transport~\cite{sundararajan_2008,burdick_2008}, drug delivery~\cite{fusco_2014,li_2017,luo_2018,sonntag_2019}, analytical sensing in biological media~\cite{balasubramanian_2010,campuzano_2011}, waste-water treatment~\cite{soler_2013}.
The propulsion mechanism of these microdevices may be classified into two generic types: external and autonomous~\cite{nadal_2014}. 
In the first type, an external field is used to propel and direct the swimmer, while in the second, the swimmer itself converts energy to achieve self--propulsion.
Deformable microswimmers, which generate propulsion by a non--reciprocal periodic deformation belong to the latter type.
One of the simplest examples is the three-linked-spheres (TLS), a swimmer built upon three spheres linked by two arms that contracts asynchronously, originally proposed by Najafi and Golestanian~\cite{najafi_2004}. 
The simplicity of this swimmer allows an analytical (within certain
approximations) and numerical study of its dynamics, making it an excellent
choice to test different numerical approaches. Experimental realizations of
the TLS have been also reported~\cite{leoni_2009,grosjean_2016,box_2017,Elder_2021}. 
In particular, the analytical and numerical studies concerning the dynamics
and optimization of the
TLS~\cite{najafi_2004,felderhof_2006,pooley_2007,golestanian_2008,alouges_2008,alexander_2008,alexander_2009,alouges_2009,vladimirov_2013,felderhof_2014,montino_2015,montino_2017,wang_2019},
are strictly valid in 
the limit where the distances between the spheres are much larger than the
sphere radius, due to the treatment of the hydrodynamic interactions.
The works of Earl \textit{et al.}~\cite{earl_2007}, and more recently Nasouri
\textit{et al.}~\cite{nasouri_2019}, Pickl and
coworkers~\cite{pickl_2012,pickl_2017} and Lengler  model the hydrodynamic interactions in more detail by means of lattice Boltzmann simulations~\cite{earl_2007,pickl_2012,pickl_2017}, multiparticle collision dynamics~\cite{earl_2007}, boundary element method~\cite{nasouri_2019} and the method of regularized Stokeslets~\cite{nunes_2021}.

In this work, we extend the theoretical study of the TLS, incorporating a much better description of the hydrodynamic interactions between the spheres composing the swimmer.
For this purpose, we use Stokesian dynamics
simulations~\cite{brady_1988,swan_2011} to study in detail the forces acting
on each of the swimmer's components and the power consumption during its
motion.
Stokesian dynamics simulations provide an accurate method to study the
dynamics of the TLS and is computationally less demanding in comparison with
mesoscale methods like lattice Boltzmann and multiparticle collision dynamics,
which consider explicitly the suspending liquid.

We define efficiency as the ratio between power dissipation and the work needed to produce the same motion by an external force. We found that the most efficient swimmer is that, in which its arms contracts almost to contact of the spheres. 
Interestingly, under these optimum conditions, the analytical predictions based on the point force (PF) approximations of the hydrodynamic equations divert significantly from those found in our simulations, in which near field interactions are taken into account. This highlights the importance of considering the finite size of the spheres, as it is done by the method implemented here. 
We believe that the results shown in this work might be very useful for the design of artificial swimmers of this kind.

The article is organized as follows: in Section~\ref{the_model}, the TLS model is presented, summarizing the point force approximation results and introducing the Stokesian dynamics approach.
Section~\ref{results} contains the results from our systematic study of the dynamics, power consumption and efficiency of the TLS. 
Finally, summary and conclusions are presented in Section~\ref{conclusions}.

\section{The model}
\label{the_model}

\begin{figure} 
\centerline{\includegraphics*[width=\figurewidth]{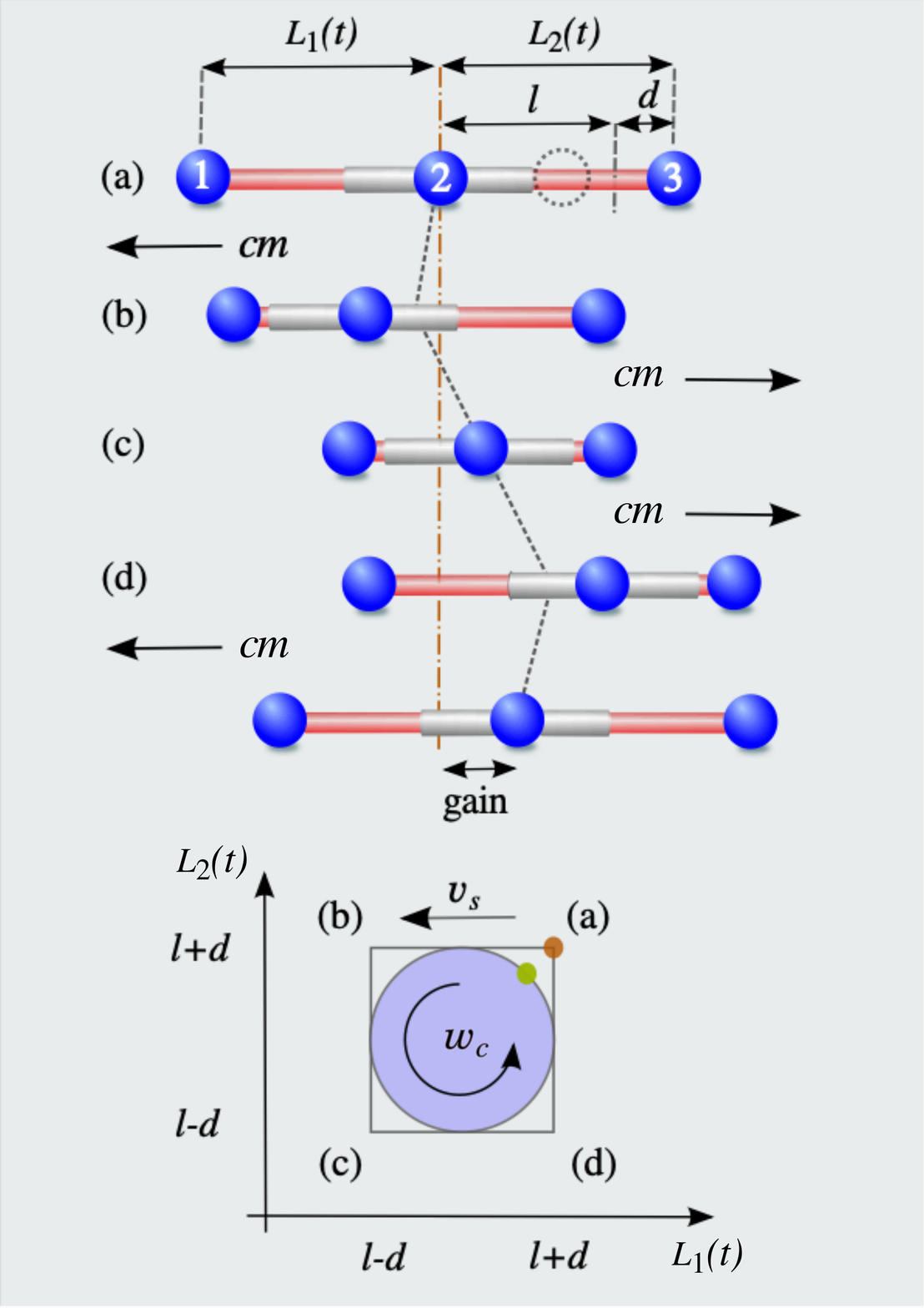}}

\caption{ Geometry of the three linked spheres swimmer. (a-d) represent the configuration changes of the swimmer during a square cycle, SC, starting from a stretched initial state. Dash dotted line corresponds to $x=0$ and dashed line corresponds to the evolution of the center of mass of the swimmer, $cm$, during the cycle. Two cycles, square and circular, are represented in the $L_1 - L_2$ space.}

\label{fig:f1}
\end{figure}

The three linked spheres swimmer (TLS) geometry is shown in Fig.~\ref{fig:f1}. It consists of three equal spheres linked by two virtual arms of lengths $L_1$ and $L_2$. 
The length of the arms varies between its contracted and its stretched states, with lengths $l_j-d$ and $l_j+d$, respectively ($j=1,2$).
$l_j$ is the arm rest length, and $d$ the amplitude of the spheres relative movement.
The swimmer stroke may be any closed cycle in the $L_1 - L_2$ phase--space. In this work, we study two particular {\it strokes}: the square cycle (SC) and the circular cycle (CC).
For the SC cycle, the stroke is defined by a square in the  $L_1 - L_2$ phase--space, that is traveled by the system at a constant speed, while for the CC the stroke is defined by a circle in the  $L_1 - L_2$ phase--space that is traveled at a constant angular velocity (see Fig.~\ref{fig:f1}).
The most remarkable difference between these two cycles is that for the SC the arms stretch/contract sequentially and at a constant speed, and in the case of the CC, while one arm stretches, the other contracts,
in a harmonic way.

\subsection{Point Force approximation}

Following the work of Golestanian and Ajdari~\cite{golestanian_2008} we write the relation between the forces $f_i$ that each sphere of radius $a$ produces on the fluid and the spheres velocities $v_i$, assuming that the spheres act like point forces on the fluid. 
Under this assumptions, which is a good approximation if $a/(l-d) \ll 1$, we have
\begin{equation}
  \begin{split}
    \pi\eta v_1 &= {f_1 \over {6a}} + {f_2 \over {4 L_1}} + {f_3 \over {4 (L_1 + L_2)}} \\ 
    \pi\eta v_2 &= {f_1 \over {4 L_1}} + {f_2 \over {6 a}} + {f_3 \over {4 L_2}} \\  
    \pi\eta v_3 &= {f_1 \over {4 (L_1 + L_2)}} + {f_2 \over {4 L_2}} + {f_3 \over {6 a}} . \\
  \end{split} 
\label{eq:v_i}
\end{equation}
\noindent
Here, $\eta$ is the fluid viscosity.
Using the self propulsion condition $f_1 + f_2 + f_3 = 0$ and defining the arms contraction velocities as $\dot{L_1} = v_2 - v_1$ and $\dot{L_2} = v_3 - v_2$,  follows:
\begin{equation}
 \pi\eta \left[ \begin{array}{c} \dot{L_1} \\ \dot{L_2} \end{array} \right] = \begin{bmatrix} A & B \\ -B & C \end{bmatrix} \times \left[ \begin{array}{c} f_1 \\ f_3 \end{array} \right],
 \label{eq:matrix}
\end{equation}
\noindent
where
\begin{equation}
  \begin{split}
    A &= {1 \over {2L_1}} - {1 \over {3a}} \\ 
    B &= {1 \over {4L_1}} + {1 \over {4L_2}} - {1 \over {4(L_1 + L_2)}} - {1 \over {6a}} \\  
    C &= {1 \over {3a}} - {1 \over {2L_2}} . \\ 
  \end{split} 
\label{eq:abc}
\end{equation}
\noindent
Which, after defining $D = AC + B^2$, leads to
\begin{equation}
  \begin{split}
    f_1 &= {{\pi\eta} \over D}(C\dot{L_1} - B\dot{L_2}) \\ 
    f_2 &= {{\pi\eta} \over D}(-(B+C)\dot{L_1} + (B-A)\dot{L_2}) \\ 
    f_3 &= {{\pi\eta} \over D}(B\dot{L_1} + A\dot{L_2}) . \\ 
  \end{split} 
\label{eq:f_i}
\end{equation}
\noindent
Equations~(\ref{eq:v_i}) and~(\ref{eq:f_i}) allows to calculate any dynamical quantity of interest, given that  $L_1$, $L_2$, $\dot{L_1}$ and $\dot{L_2}$ are known as a function of time, $t$, {\it i.e.} the particular stroke cycle is determined. 
For the circular cycle, the arms deformations are given by
\begin{equation}
  \begin{split}
    L_1(t) &= l + d \cos(w_c t + \pi/4) \\ 
    L_2(t) &= l + d \sin(w_c t + \pi/4) \\ 
  \end{split} 
\label{eq:CC}
\end{equation}
\noindent
with the angular velocity, $w_c$, and a corresponding period, $T_c = 2\pi/w_c$. 
For the square cycle,
\begin{equation}
  \begin{cases}
  L_1(t) = l + d - v_st,& L_2(t) = l + d ,  \hspace{11.3mm}      t \in I_1 \\ 
  L_1(t) = l - d ,      & L_2(t) = l + d - v_st, \hspace{2mm}  t \in I_2 \\ 
  L_1(t) = l - d + v_st,& L_2(t) = l - d ,  \hspace{11.3mm}      t \in I_3 \\
  L_1(t) = l + d ,      & L_2(t) = l - d + v_st , \hspace{2mm} t \in I_4 \\ 
  \end{cases}
\label{eq:SC}
\end{equation}
\noindent
Here,  $v_s$ is the contraction velocity of the arms, the period of the motion is given by $T_s = 8d/v_s$ and $I_i$ are consecutive intervals of duration $T_s/4$.

\subsection{Stokesian Dynamics}

To take into account the full hydrodynamic interactions between the spheres composing the TLS, one could solve the full three--body problem.
This, however, constitutes a formidable task.
For this reason, and considering that the interactions between two TLS swimmers or even a suspension of TLS swimmers might be of interest, we study the dynamic of the swimmer by implementing Stokesian Dynamics~\cite{brady_1988} (SD) simulations. 
This simulation scheme has been extended to treat self--propulsion~\cite{swan_2011}, as long as the swimmer can be approximated by a collection of spheres, which in the particular case of the TLS swimmer is not even an approximation. 
Stokesian dynamics is a well--established simulation scheme for the study of the suspensions taking into account the many--body hydrodynamic interactions. 
It has been shown that it can quantitatively reproduce the properties of
monodisperse suspensions at high volume fractions~\cite{bossis_1988} and has
been successfully applied to study the dynamics and rheology of colloidal
particles~\cite{foss_2000, phung_1996,heinen_2010, heinen_2011, banchio_2018}.

Here, we have implemented SD simulations adapting the code provided in the
work of Swan and coworkers~\cite{swan_2011} to represent the TLS swimmer with
the two swimming strokes cycles under consideration, namely, the square cycle
and the circular cycle.

For convenience, we use the sphere radius, $a$, as unit length, the cycle period, $T$, as unit of time, and $\eta a^2/T$ as unit of force, where $\eta$ represents the fluid viscosity.
The time step used in the SD simulations was $dt = T/n_0$, with $n_0$ typically of the order of 50000, in the case of the CC cycle, while for the SC cycle, $n_0$ was 400000. The SC cycle needs to be solved with a much smaller time step due to the fact that the spheres come almost to contact at a constant velocity.

\section{RESULTS}
\label{results}
\subsection{Swimmer Dynamics}
\begin{figure*}[t] 
\centerline{\includegraphics*[width=0.85\textwidth]{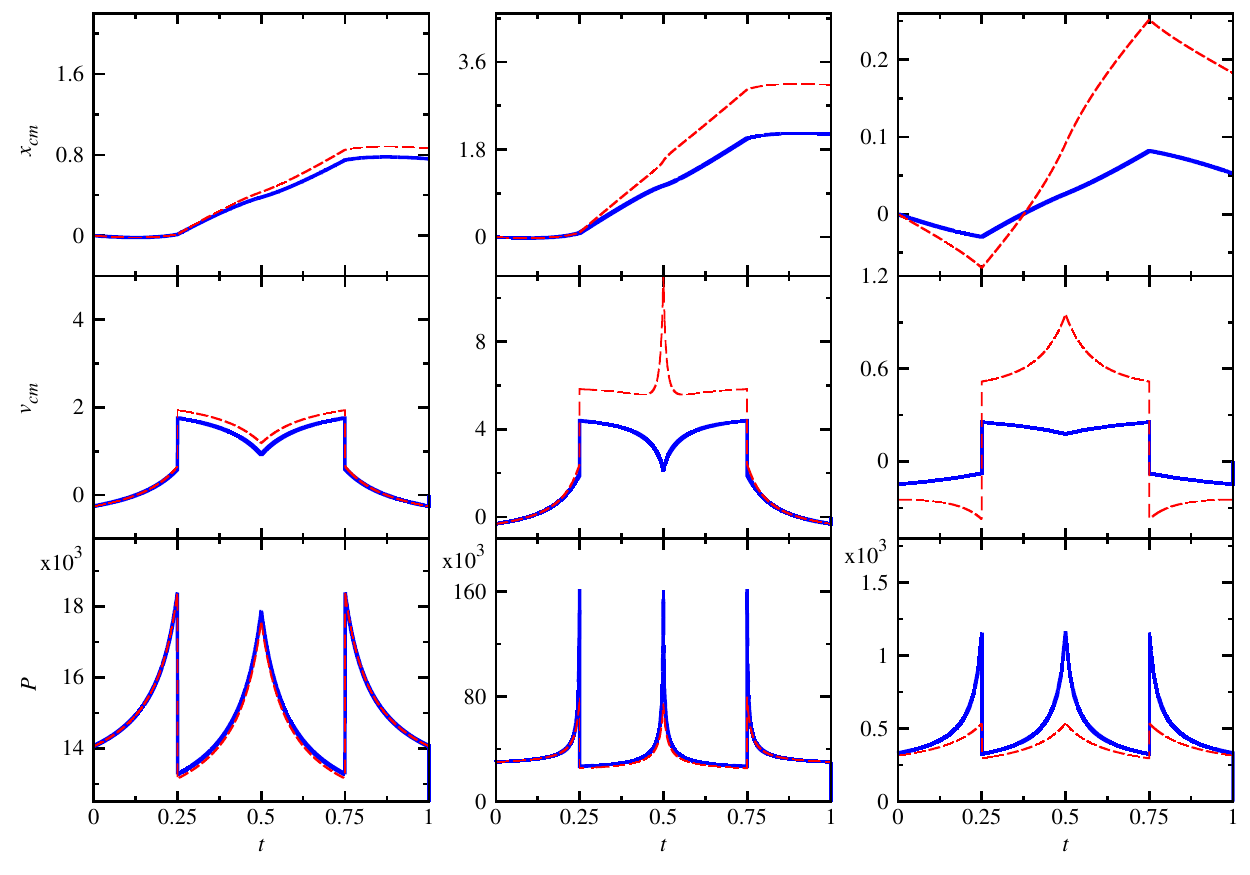}}
\caption{Comparing Stokesian Dynamics and the point force approximation:
  square cycle swimmers. The
  center of mass position, $x_{cm}$, center of mass velocity, $v_{cm}$, and
  the dissipated power, $P$, as function of time, $t$, for 3 different square cycle
  swimmers.
  From left to right: Column 1, $l$ = 8 and $d$ = 4. Column 2, $l$ = 8 and $d$
  = 5.9. Column 3 $l$ = 2.6 and $d$ = 0.5. In all panels solid blue lines
  corresponds to Stokesian Dynamics results and dashed red lines to the point
  force approximation. }
\label{fig:f2}
\end{figure*}
Three linked sphere swimmer takes advantage of the differences in the drag forces on the different intervals of its swimming cycle to produce its net displacement. 
If the cycle in Fig.~\ref{fig:f1} is traveled counterclockwise, the swimmer moves to the right, and if the cycle is traveled clockwise the swimmer moves to the left. 
In the first part of the SC cycle, starting with both arms extended and going through the cycle counterclockwise, the left arm contracts at a constant speed $v_s$ from (a) $L_1 = l + d$  to (b) $L_1 = l -d$. 
This contraction moves the center of mass, $cm$, to the left as it is shown by the gray dashed line. 
During this interval, the other arm is extended, resulting in a larger drag opposing the backward motion of the swimmer. In the next interval, $L_2$  goes from $l+d$ to $l-d$, producing a $cm$ motion to the right. 
This forward movement is larger than the previous displacement to the left due to the lower drag exerted by the left arm, that is contracted. 
Analogously, analyzing the rest of the cycle, a net displacement (gain) of the swimmer to the right is obtained.

\begin{figure*}[t] 
\centerline{\includegraphics*[width=0.85\textwidth]{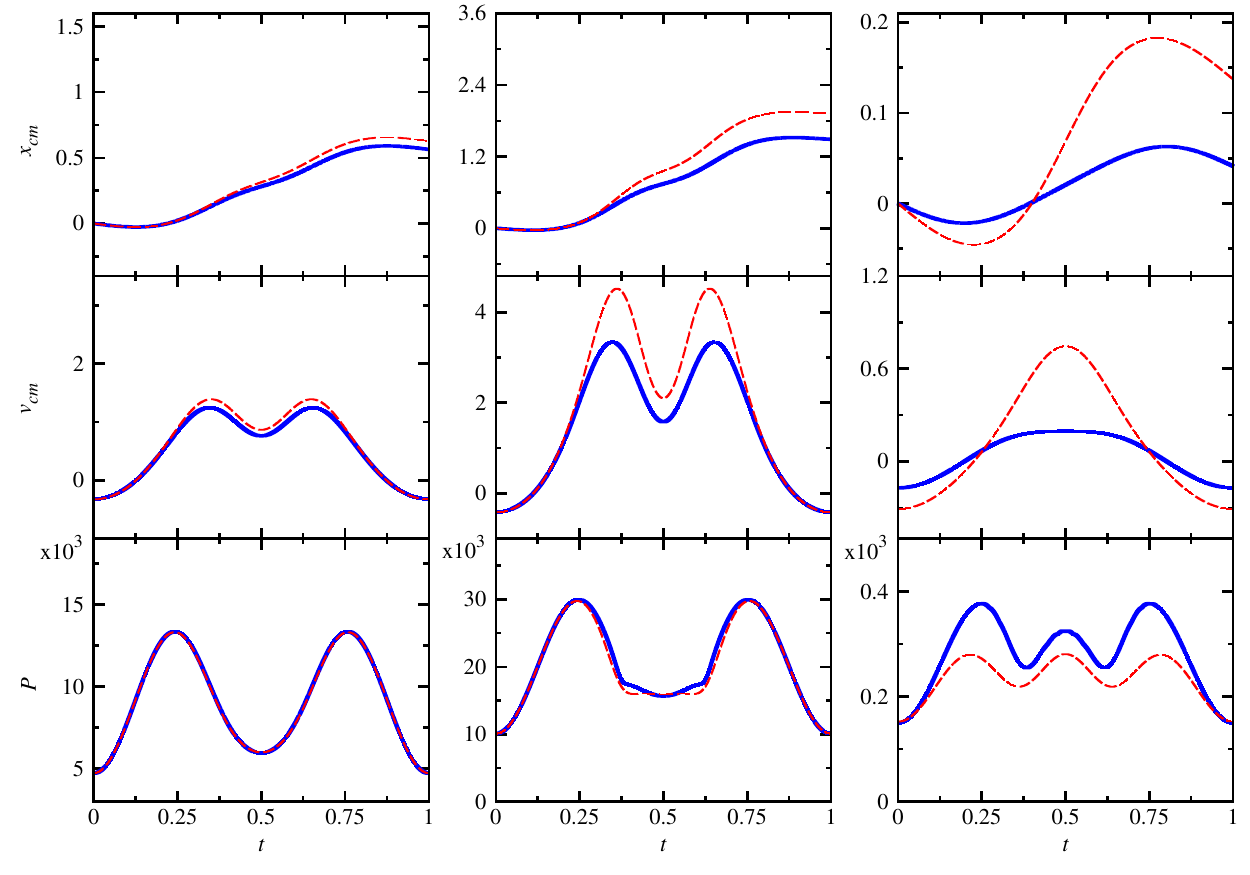}}
\caption{Comparing Stokesian Dynamics and the point force approximation:
  circular cycle swimmers. The
  center of mass position, $x_{cm}$, center of mass velocity, $v_{cm}$, and
  the dissipated power, $P$, as function of time, $t$, for 3 different circular cycle
  swimmers.
  From left to right: Column 1, $l$ = 8 and $d$ = 4. Column 2, $l$ = 8 and $d$
  = 5.9. Column 3 $l$ = 2.6 and $d$ = 0.5. In all panels solid blue lines
  corresponds to Stokesian Dynamics results and dashed red lines to the point
  force approximation. }
\label{fig:f3}
\end{figure*}

The situation, depicted in the last paragraph, is shown quantitatively in
Fig.~\ref{fig:f2} for three swimmers that differ in their stretched--arm
sizes, $l+d$, and in the amplitude of their arms motion, $d$. Blue solid lines
are results obtained by Stokesian Dynamics, while dashed red lines correspond
to the point force approximations. 
The first column shows results for a swimmer, $s_1$, with $l=8$ and $d=4$.
With these dimensions, the closest distance between sphere centers, $l-d = 4$,
is large enough to allow for a fine estimation of the swimmer dynamics by the
point force approximation. 
The second column corresponds to $s_2$, a swimmer also with $l=8$, but with
$d=5.9$.
Note that for this swimmer, the extreme spheres do come almost to contact (actually, at a surface separation of $0.1$) with the central sphere when the respective arm is fully contracted. 
Finally, the third column display results for a smaller swimmer, $s_3$, in which the spheres are almost all the time close to contact ($l=2.6$, $d=0.5$).
Figure.~\ref{fig:f3} shows analogous results for the same swimmers, but following a circular cycle.

For the three swimmers, the backward motion during the first interval can be seen both in the time evolution of the center of mass position and in an initially negative center of mass velocity. 
Looking at the final displacement after one cycle, it is observed that $s_2$ is the swimmer that moves more in one cycle (this is also true for the circular cycle shown in Fig.~\ref{fig:f3}). 
The difference between the SD results and PF approximations is significant for
all the variables and grows with the compression of the arms, as expected,
showing even a curvature inversion for the velocity of swimmers $s_2$ and
$s_3$ in the square cycle.
This inversion takes place at the regions where the spheres get in close proximity and the need for a complete representation of the hydrodynamic forces is more relevant. 

\begin{figure} 
\centerline{\includegraphics*[width=\figurewidth]{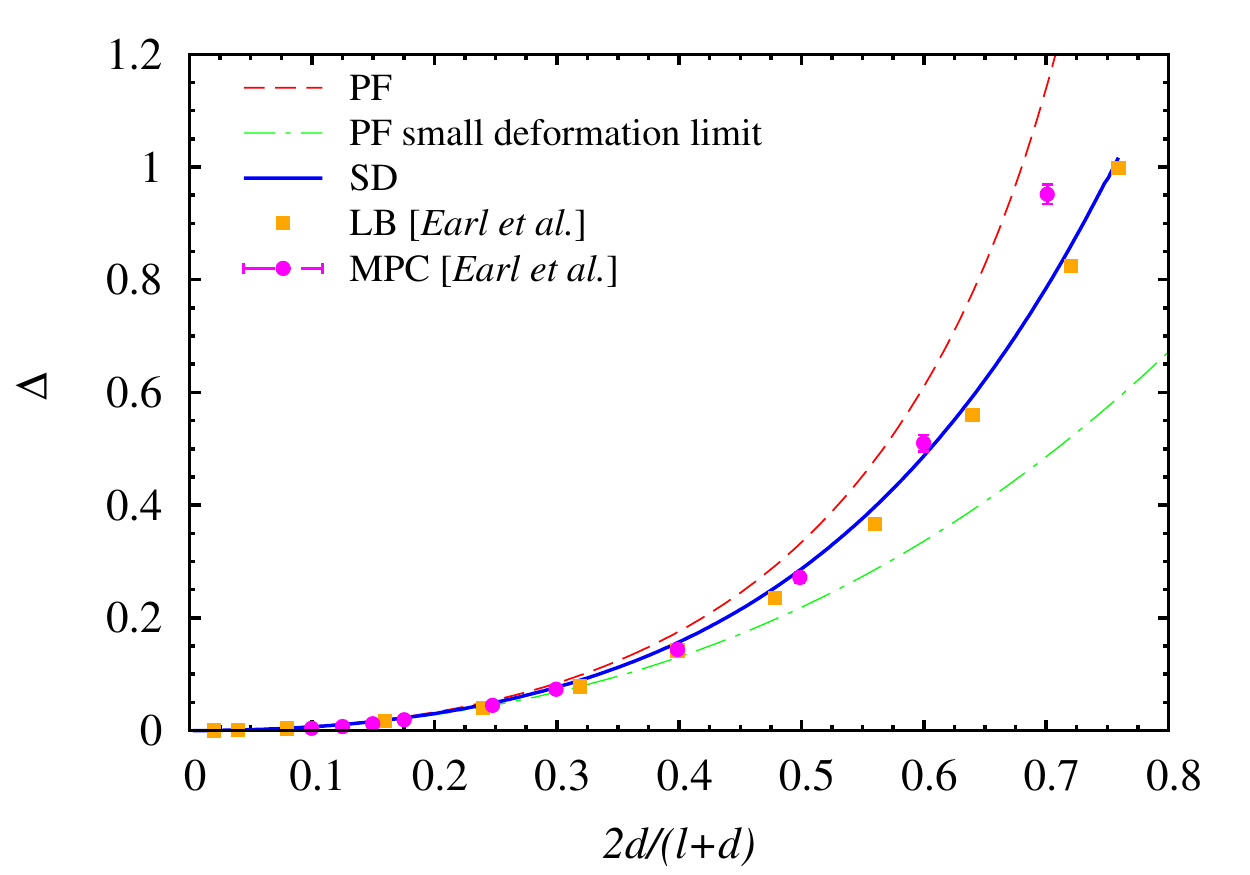}}

\caption{Net displacement in one cycle, $\Delta$, 
for the SC swimmer as a function of the total spheres relative
displacement ($2d$) divide by the maximum arm length ($l+d$). For
the PF (dashed red line) and SD (solid blue line) calculations 
$l+d=25/3$ was used, for consistency with Lattice Boltzman (orange-squares) and MPC
(pink-circles) results taken from Ref.~\onlinecite{earl_2007}.
The small deformation limit within the PF approximation (dash--dotted green line) was
obtained from Eq.~(22) in Ref.~\onlinecite{earl_2007}.
}

\label{fig:f3_5}
\end{figure}

To compare SD results with other methods that also include hydrodynamic interactions, like Lattice Boltzmann (LB)~\cite{ladd_1994a,ladd_1994b} and Multiparticle Collisions Dynamics (MPC)~\cite{malevanets_1999}, we have calculated the one cycle net displacement, $\Delta$, for the SC swimmer with $l+d=25/3$ as a function of the relative spheres displacement amplitude, and compared SD results with those obtained by Earl \textit{et al.}~\cite{earl_2007}.
This comparison in conjunction with analytical results obtained within the
point force approximation are shown in Figure~\ref{fig:f3_5}.
LB and MPC data have been taken from the work of Earl \textit{et al.}~\cite{earl_2007}, where they use both mesoscale methods to study the TLS and other generalizations of it. For details in the implementations of LB and MPC and the parameters used, see Ref.~\onlinecite{earl_2007}.
Remarkably, the SD results are in quite good agreement with both methods, and
in particular with LB, which is more accurate.
Note, however, that the LB implementation in Ref.~\onlinecite{earl_2007} do
not include lubrication corrections~\cite{nguyen_2002}, for this reason near field interparticle
interactions might be underestimated when particles come close together.
The great advantage of SD, in comparison with those mesoscale schemes, is that
it treats the fluid as a continuum, allowing the simulation of larger time-scales as well as larger systems with much less computational effort.
Figure~\ref{fig:f3_5} also includes the small deformation limit of the net
displacement within the PF approximation up to second order in $2d/(l+d)$,
Eq.~(22) from Ref.~\onlinecite{earl_2007}. As expected, in the limit of small
deformation it converges to the full PF approximation.

The instantaneous dissipated power can be obtained directly from the expression
\begin{eqnarray}
P(t) &=& f_1 v_1 + f_2 v_2 + f_3 v_3 \\
&=& -f_1 \dot{L_1} + f_3 \dot{L_2} .
\end{eqnarray}
\noindent
Note that with the adimensionalization we are using, $P(t)$ is expressed in units of $\eta a^3 / T^2$.

A remarkable increase of the dissipation is found for swimmers $s_2$ and $s_3$ in the square cycle with respect to $s_1$. This compression dependent behavior is highly underestimated by the PF approximation and will have a major influence in the determination of the swimmer efficiency. 
The fast growth of the dissipation in the square cycle is caused by the spheres approximating at an imposed constant speed $v_s$, overcoming the lubrication forces growing like $1/(L_i-2)$. These sharp peaks are not present for the circular cycle because, in this case, the contraction velocity of the arms goes to zero when the spheres are at the shortest separation.

\subsection{Averages quantities}

\begin{figure} 
\centerline{\includegraphics*[width=\figurewidth]{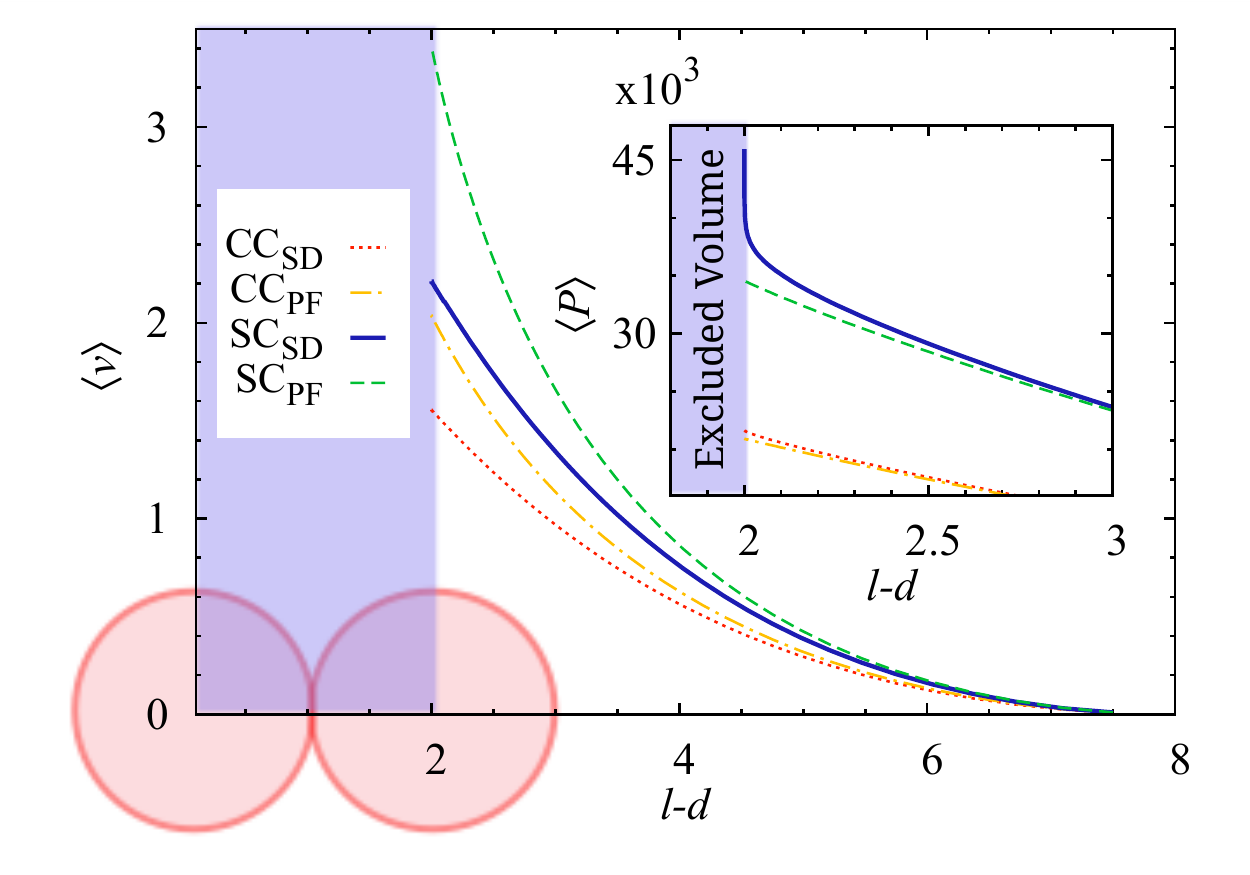}}
\caption{Average velocity and power dissipation (inset) for swimmers with
  $l=8$ as a function of
  of the contracted--arm length, $l-d$. The shaded area represent volume exclusion due to the finite size of the spheres. 
Dotted red  and dash--dotted yellow lines correspond to results for the CC swimmer obtained by SD and PF respectively. 
Solid blue and dashed green lines correspond to results for the SC swimmer obtained by SD and PF respectively. }
\label{fig:f4}
\end{figure}

\begin{figure} 
\centerline{\includegraphics*[width=\figurewidth]{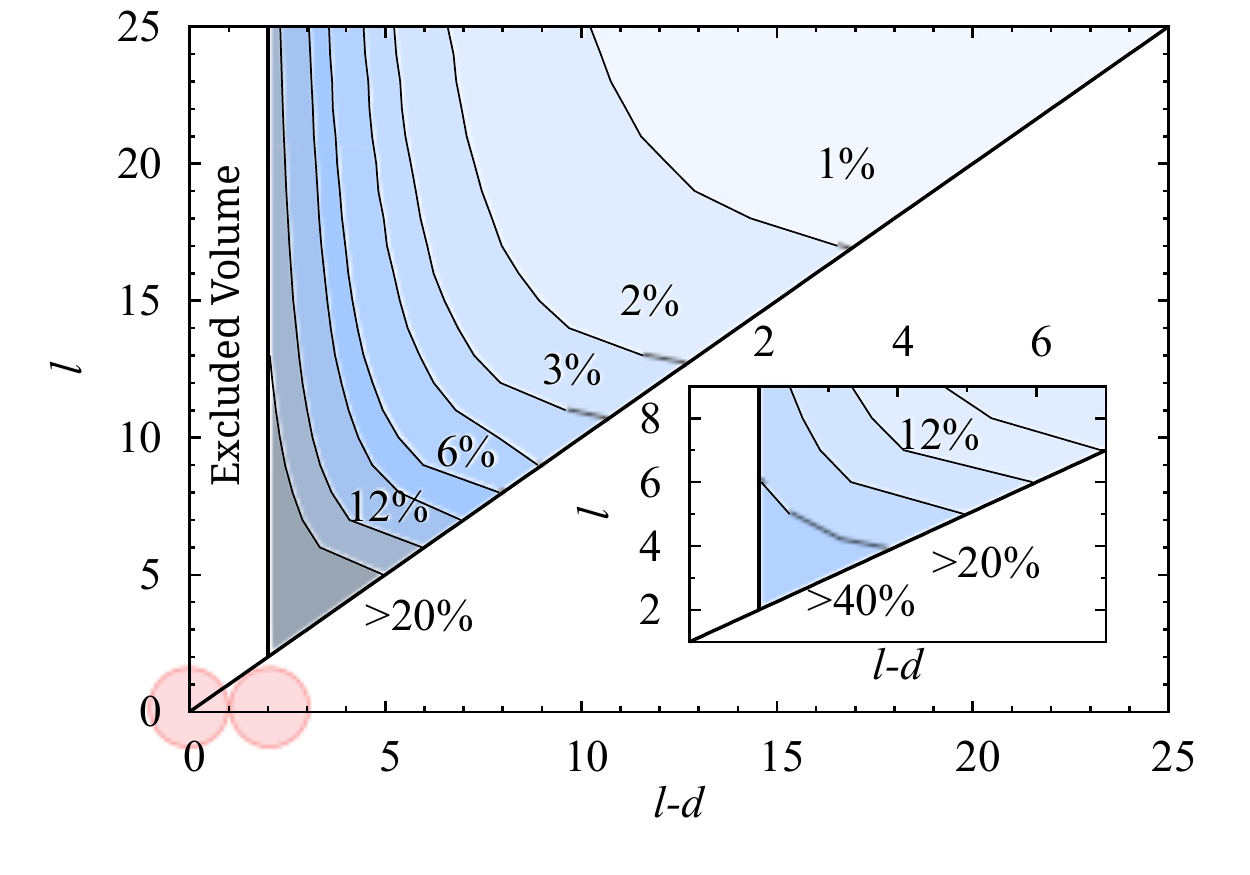}}
\caption{Percentage {\it error} (difference) map of  $\langle v \rangle$ obtained by PF with respect to SD, for a CC swimmer. The triangular domain is defined by $l-d \geq 2$ required by volume exclusion and $d \geq 0$.}
\label{fig:f5}
\end{figure}

To precisely quantify the differences between the PF approximation and the SD results, we have computed the mean velocity, $\langle v \rangle$, and the mean dissipated power, $\langle P \rangle$, averaged over one period. With the selected unit of time, the mean velocity and the mean dissipated power are equivalent to the net displacement, $\Delta$, and the dissipated power per cycle, respectively. 
Fig.~\ref{fig:f4} shows these quantities 
as a function of the minimum distance between the sphere centers, $l-d$. 
As can be seen in Fig.~\ref{fig:f4} for a swimmer with $l=8$, the PF approximation overestimates the velocity found by SD, both for SC and CC, and underestimates the average power dissipation (inset of Fig.~\ref{fig:f4}). 
We have also analyzed the percentage {\it error} (difference) between the PF
and SD mean velocities, $100(\langle v_{PF} \rangle - \langle v_{SD}
\rangle)/\langle v_{SD} \rangle$, in terms of the rest length of the arms, $l$, and their highest compression $l-d$. 
These results are summarized in Fig.~\ref{fig:f5}, where the corresponding percentage--error map is shown. Note that a value of $l > 17$ is required for a swimmer to obtain an error smaller than 1$\%$ 
when using the PF approximation. 
Even for a swimmer with $l=20$, an error $<$1$\%$ is obtained for extremely low compression, which are only possible for amplitudes larger than $d\approx 5$. 
On the other hand, already for minimum separation as large as 4 the error may be larger that 20\%.
Summarizing, the PF approximation is only good for swimmers with large arm lengths, $l$, and, simultaneously, large minimum sphere separations, $l-d$, compared with the sphere radius (roughly, $l-d$ larger than 15 for an error $<$1$\%$). 

\subsection{Efficiency}

For studying the efficiency, $\varepsilon$, of the swimmer we use a definition first introduced by Lighthill~\cite{lighthill_1976}, which corresponds to the ratio between the power dissipated when an external force moves the swimmer at a given velocity, and the power dissipated by the swimmer to propel itself at the same velocity.
In the particular case of the TLS, its shape, and consequently the drag force, varies during a stroke cycle. For this reason, we calculate the efficiency as
\begin{equation}
\varepsilon = C(l,d)\, \langle v \rangle^2 / \langle P \rangle, 
\end{equation}
\noindent
where $C(l,d)$ is the friction coefficient of a non-deforming TLS swimmer with
arms rest length, $l$, and arm variation length amplitude, $d$, in its most contracted state ($L_1 = L_2 = l-d$). 
With this choice, we calculate a lower limit for the efficiency.
Other authors~\cite{bet_2017} suggest using an average friction coefficient, corresponding to the time evolution of the swimmer shape during the stroke. We prefer to use the less dissipative configuration (the most contracted) to define the efficiency since the shape changes are consequence of the swimming stroke.
The coefficient $C(l,d)$ was obtained within the same SD simulation scheme, and takes values between $(0.51\pm 0.01)\times 18\pi$ and $18\pi$ in units of $\eta a$, corresponding to the limits of three spheres in contact  ($l=2$ with $d=0$) and infinitely separated, respectively.

\begin{figure}[htb]
\centerline{\includegraphics*[width=\figurewidth]{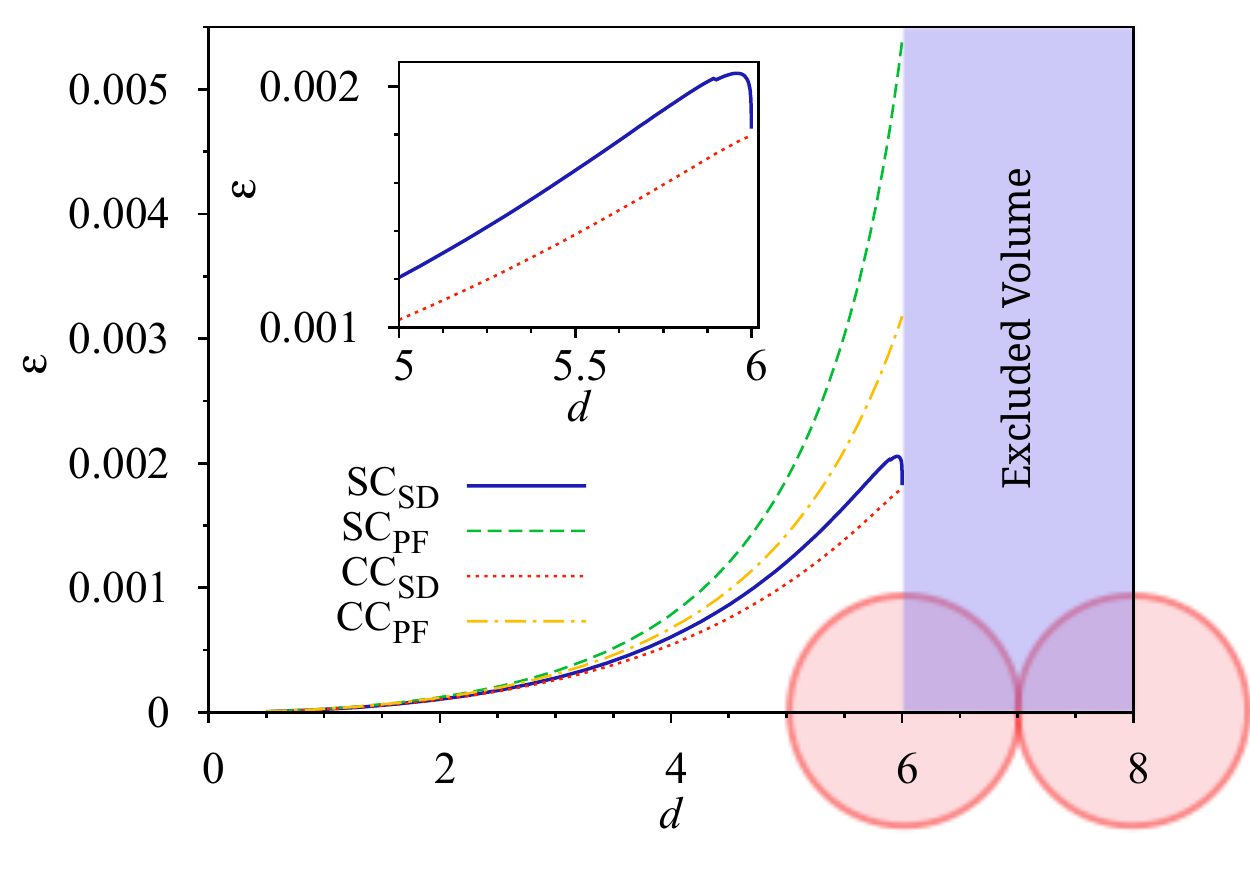}}
\caption{Efficiencies of a swimmer with $l=8$ as a function of the amplitude, $d$. Solid blue line and dotted red line correspond to SC and CC respectively, calculated using SD. Dashed green line and dot--dashed yellow line shows the overestimation of the efficiency for the square and circular cycles respectively calculated using PF approximation. 
The shaded area represents volume exclusion. Inset: different behavior of the SC and CC for spheres close to contact.}
\label{fig:f6}
\end{figure}

As it was shown in Fig.~\ref{fig:f4} for a swimmer with arm length $l=8$, both cycle types present an average velocity, $\langle v \rangle$, and {an average dissipated power,} $\langle P \rangle$, that grow  when the minimum sphere separation (contracted arm length), $l-d$, decreases, reaching a maximum when the spheres can touch.
The corresponding point force approximation results are also shown for comparison. The PF starts to overestimate (appreciably) the average velocity and dissipated power for both systems when the contracted arm length is around 6.
The faster growth of $\langle v \rangle^2$ in relation to that of $\langle P \rangle$ produces also an increasing efficiency for a larger $d$, as can be seen in Fig.~\ref{fig:f6} for both cycles. 
This growing of the efficiency stops when $d$ is large enough for the outermost spheres to approximate to the central sphere and almost touch it for a SC swimmer, and when $d=l-2$ for a CC swimmer. 
In the case of SC swimmer,  the  lubrication forces produces a sharp increase in power dissipation and a corresponding fall of the efficiency. This behavior of the efficiency can not be found using the PF approximation, because it does not account for these forces, and for this reason it produces the most significant overestimation of the efficiency  where the swimmers are more efficient. 
In the inset of Fig.~\ref{fig:f6} it is possible to see that the lubrication forces affect the efficiency of the two cycles in markedly different ways. 
For the square cycle, the efficiency reaches a maximum for swimmers that have a contracted arm length ($l-d$) around 2.1, i.e. a gap between spheres approximately $10\%$ of the sphere radius. For the circular cycle, on the other hand, no maximum is observed, and the efficiency grows monotonically as the contracted swimmer size decreases. These different behaviors are produced by the different relative velocities of the spheres when approaching each other. For the square cycle, the approximation velocity is constant and equal to $v_s$, while for the circular cycle it goes to zero as a sine function.

\begin{figure}[htb] 
\centerline{\includegraphics*[width=\figurewidth]{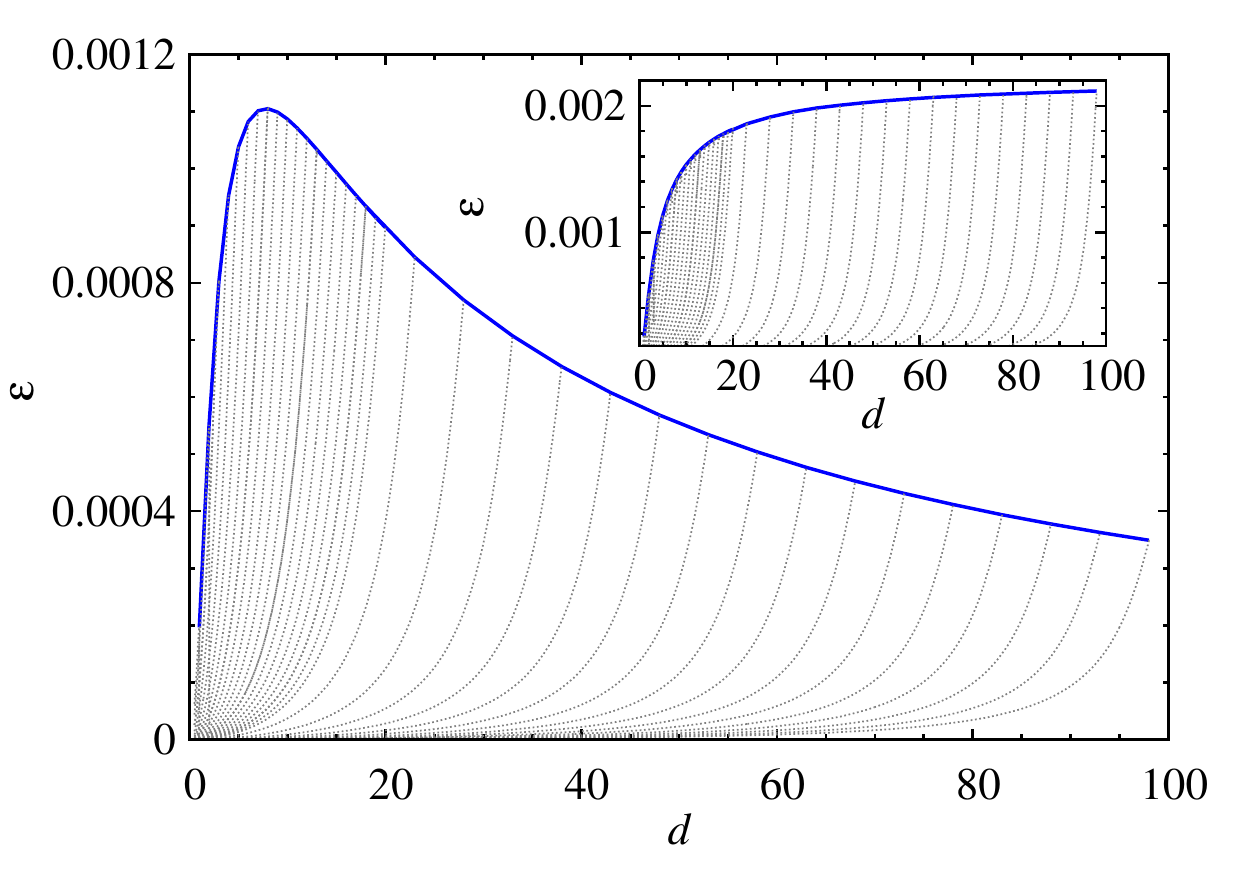}}
\caption{Efficiency of the circular cycle swimmer for different arm rest
  length, $l$, as a function of amplitude, $d$, (thin gray
  lines) and efficiency of the most efficient swimmer for each $l$ as a function of $d$
  (thick blue lines). The inset shows the corresponding results for the square cycle swimmer.}
\label{fig:f7}
\end{figure}

\begin{figure}[t!] 
\centerline{\includegraphics*[width=\figurewidth]{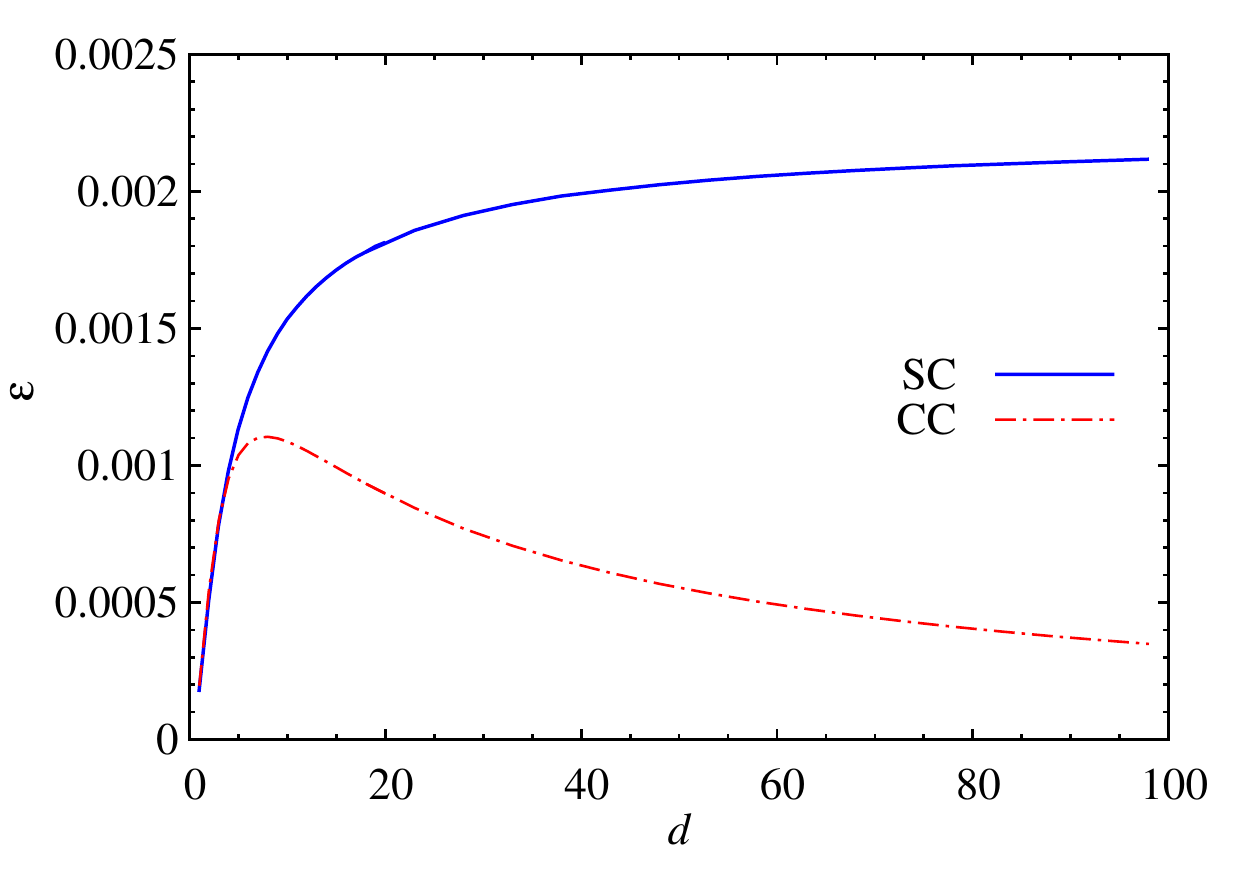}}
\caption{Comparison of the most efficient swimmer for the square cycle (blue solid line) with the most efficient swimmer for the circular cycle (red dot--dashed line) as a function of the amplitude, $d$.}
\label{fig:f10}
\end{figure}

The efficiencies as a function of the amplitude, $d$, for different values of the arm length, $l$, are shown in Fig.~\ref{fig:f7} as thin gray lines. 
Note that for a given value of $l$, $d$ can take values between $0$ and $l-2$, and that we have plotted here only the region where the efficiency grows with $d$, truncating the (gray) lines when they reach the respective maximum efficiency.
Connecting the highest efficiency points for each $l$ it is possible to build a curve of the maximum efficiency of the swimmer as a function of $d$ (for each $d$ there is a system with $l\sim d+2$ that has the maximum possible efficiency). 
This curve is interesting because it allows to visualize how the swimmer size does affects its ability to swim. In Fig.~\ref{fig:f10} a comparison for this function for both, SC and CC cycle swimmers is presented. Notably, the two cycles studied present markedly different behaviors. In the case of the circular cycle, the efficiency of the best swimmer with a given $l$ grows almost linearly for small swimmers, then has its maximum at $d=8$ and finally decays monotonically for larger swimmers. 
The Square Cycle, on the other hand, does not show an optimum size and has an efficiency, $\epsilon$, that monotonically grows and tends asymptotically to a value slightly over 0.0021. 
Due to these two distinct behaviors of the most efficient swimmers, it would be preferable to build large swimmers using a square cycle rather than a circular cycle, but in the case of small swimmers the specific shape of the cycle seems to be less important.

\section{Conclusions}
\label{conclusions}

We have systematically studied by implementing SD simulations, the dynamics of the TLS swimmer for a wide range of parameters and two different swimming cycles, and compared the results with the analytic point force approximation.
Furthermore, the efficiency of the swimmer was analyzed and the optimum parameters were identified.
The point force approximation describes reasonably well the dynamics, as long as the spheres do not come into close contact, where the strong lubrication forces start to play a dominant role. 
This has been quantitatively analyzed, and the results presented in a figure showing the percentage difference in the mean velocity between point force approximation and the SD solution, as a function of the size of the arms and the minimum separation of the spheres.

Our study has shown that the velocity and the dissipated power grows for a given size of the arms, $l$, when the minimum separation, $l-d$, decreases.
Furthermore, the mean velocity and the mean dissipated power in one cycle, are larger for the SC cycle, for any value of $d$.
This result alone is not sufficient to decide which cycle is most efficient since a larger velocity has a larger energetic cost.

Considering the most efficient swimmer for a given amplitude, $d$, we have shown that the two studied swimming cycles have nearly the same efficiency as long as $d \lesssim 8$, and that for
larger separations (i.e. also larger swimmers) the SC cycle results more efficient than the CC.

Summarizing, we have shown that the SD simulation scheme is an appropriate tool to study the dynamics and efficiency of artificial swimmers, in particular those constituted by spherical particles. 
The precise description of the hydrodynamic interactions can be of relevance for boosting the study and design of more complex and efficient micro-swimmers or micro-machines with individual constitutive parts of finite size, that can come very close, positioning indeed Stokesian Dynamics as a valuable tool for these purposes. Visualizing its promising applications as natural ongoing steps we are studying interaction between two or more TLS swimmers, as well as more complex micro-swimmer models composed by hundreds of spheres.

\section{Acknowledgments}
This work was supported by SECyT (Universidad Nacional de Córdoba); CONICET; and Agencia Nacional de Promoción Científica y Tecnológica under Grant No.  PICT 2015-0735.

\section{Data Availability}
The data that support the findings of this study are available from the corresponding author upon reasonable request.

\bibliography{active_matter}

\begin{thebibliography}{54}%
\makeatletter
\providecommand \@ifxundefined [1]{%
 \@ifx{#1\undefined}
}%
\providecommand \@ifnum [1]{%
 \ifnum #1\expandafter \@firstoftwo
 \else \expandafter \@secondoftwo
 \fi
}%
\providecommand \@ifx [1]{%
 \ifx #1\expandafter \@firstoftwo
 \else \expandafter \@secondoftwo
 \fi
}%
\providecommand \natexlab [1]{#1}%
\providecommand \enquote  [1]{``#1''}%
\providecommand \bibnamefont  [1]{#1}%
\providecommand \bibfnamefont [1]{#1}%
\providecommand \citenamefont [1]{#1}%
\providecommand \href@noop [0]{\@secondoftwo}%
\providecommand \href [0]{\begingroup \@sanitize@url \@href}%
\providecommand \@href[1]{\@@startlink{#1}\@@href}%
\providecommand \@@href[1]{\endgroup#1\@@endlink}%
\providecommand \@sanitize@url [0]{\catcode `\\12\catcode `\$12\catcode
  `\&12\catcode `\#12\catcode `\^12\catcode `\_12\catcode `\%12\relax}%
\providecommand \@@startlink[1]{}%
\providecommand \@@endlink[0]{}%
\providecommand \url  [0]{\begingroup\@sanitize@url \@url }%
\providecommand \@url [1]{\endgroup\@href {#1}{\urlprefix }}%
\providecommand \urlprefix  [0]{URL }%
\providecommand \Eprint [0]{\href }%
\providecommand \doibase [0]{https://doi.org/}%
\providecommand \selectlanguage [0]{\@gobble}%
\providecommand \bibinfo  [0]{\@secondoftwo}%
\providecommand \bibfield  [0]{\@secondoftwo}%
\providecommand \translation [1]{[#1]}%
\providecommand \BibitemOpen [0]{}%
\providecommand \bibitemStop [0]{}%
\providecommand \bibitemNoStop [0]{.\EOS\space}%
\providecommand \EOS [0]{\spacefactor3000\relax}%
\providecommand \BibitemShut  [1]{\csname bibitem#1\endcsname}%
\let\auto@bib@innerbib\@empty
\bibitem [{\citenamefont {Purcell}(1977)}]{purcell_1977}%
  \BibitemOpen
  \bibfield  {author} {\bibinfo {author} {\bibfnamefont {E.~M.}\ \bibnamefont
  {Purcell}},\ }\bibfield  {title} {\bibinfo {title} {Life at low reynolds
  number},\ }\href {https://doi.org/10.1119/1.10903} {\bibfield  {journal}
  {\bibinfo  {journal} {Am. J. Phys.}\ }\textbf {\bibinfo {volume} {45}},\
  \bibinfo {pages} {3} (\bibinfo {year} {1977})}\BibitemShut {NoStop}%
\bibitem [{\citenamefont {Lauga}\ and\ \citenamefont
  {Powers}(2009)}]{lauga_2009}%
  \BibitemOpen
  \bibfield  {author} {\bibinfo {author} {\bibfnamefont {E.}~\bibnamefont
  {Lauga}}\ and\ \bibinfo {author} {\bibfnamefont {T.~R.}\ \bibnamefont
  {Powers}},\ }\bibfield  {title} {\bibinfo {title} {The hydrodynamics of
  swimming microorganisms},\ }\href
  {https://doi.org/10.1088/0034-4885/72/9/096601} {\bibfield  {journal}
  {\bibinfo  {journal} {Rep. Prog. Phys.}\ }\textbf {\bibinfo {volume} {72}},\
  \bibinfo {pages} {096601} (\bibinfo {year} {2009})}\BibitemShut {NoStop}%
\bibitem [{\citenamefont {Elgeti}\ \emph {et~al.}(2015)\citenamefont {Elgeti},
  \citenamefont {Winkler},\ and\ \citenamefont {Gompper}}]{elgeti_2015}%
  \BibitemOpen
  \bibfield  {author} {\bibinfo {author} {\bibfnamefont {J.}~\bibnamefont
  {Elgeti}}, \bibinfo {author} {\bibfnamefont {R.~G.}\ \bibnamefont
  {Winkler}},\ and\ \bibinfo {author} {\bibfnamefont {G.}~\bibnamefont
  {Gompper}},\ }\bibfield  {title} {\bibinfo {title} {Physics of
  microswimmers--single particle motion and collective behavior: a review},\
  }\href {https://doi.org/10.1088/0034-4885/78/5/056601} {\bibfield  {journal}
  {\bibinfo  {journal} {Rep. Prog. Phys.}\ }\textbf {\bibinfo {volume} {78}},\
  \bibinfo {pages} {056601} (\bibinfo {year} {2015})}\BibitemShut {NoStop}%
\bibitem [{\citenamefont {Lauga}(2016)}]{lauga_2016}%
  \BibitemOpen
  \bibfield  {author} {\bibinfo {author} {\bibfnamefont {E.}~\bibnamefont
  {Lauga}},\ }\bibfield  {title} {\bibinfo {title} {Bacterial hydrodynamics},\
  }\href {https://doi.org/10.1146/annurev-fluid-122414-034606} {\bibfield
  {journal} {\bibinfo  {journal} {Annu. Rev. Fluid Mech.}\ }\textbf {\bibinfo
  {volume} {48}},\ \bibinfo {pages} {105} (\bibinfo {year} {2016})}\BibitemShut
  {NoStop}%
\bibitem [{\citenamefont {Gompper}\ \emph {et~al.}(2020)\citenamefont
  {Gompper}, \citenamefont {Winkler}, \citenamefont {Speck}, \citenamefont
  {Solon}, \citenamefont {Nardini}, \citenamefont {Peruani}, \citenamefont
  {Löwen}, \citenamefont {Golestanian}, \citenamefont {Kaupp}, \citenamefont
  {Alvarez}, \citenamefont {Ki{\o}rboe}, \citenamefont {Lauga}, \citenamefont
  {Poon}, \citenamefont {DeSimone}, \citenamefont {Mui{\~{n}}os-Landin},
  \citenamefont {Fischer}, \citenamefont {Söker}, \citenamefont {Cichos},
  \citenamefont {Kapral}, \citenamefont {Gaspard}, \citenamefont {Ripoll},
  \citenamefont {Sagues}, \citenamefont {Doostmohammadi}, \citenamefont
  {Yeomans}, \citenamefont {Aranson}, \citenamefont {Bechinger}, \citenamefont
  {Stark}, \citenamefont {Hemelrijk}, \citenamefont {Nedelec}, \citenamefont
  {Sarkar}, \citenamefont {Aryaksama}, \citenamefont {Lacroix}, \citenamefont
  {Duclos}, \citenamefont {Yashunsky}, \citenamefont {Silberzan}, \citenamefont
  {Arroyo},\ and\ \citenamefont {Kale}}]{Gompper_2020}%
  \BibitemOpen
  \bibfield  {author} {\bibinfo {author} {\bibfnamefont {G.}~\bibnamefont
  {Gompper}}, \bibinfo {author} {\bibfnamefont {R.~G.}\ \bibnamefont
  {Winkler}}, \bibinfo {author} {\bibfnamefont {T.}~\bibnamefont {Speck}},
  \bibinfo {author} {\bibfnamefont {A.}~\bibnamefont {Solon}}, \bibinfo
  {author} {\bibfnamefont {C.}~\bibnamefont {Nardini}}, \bibinfo {author}
  {\bibfnamefont {F.}~\bibnamefont {Peruani}}, \bibinfo {author} {\bibfnamefont
  {H.}~\bibnamefont {Löwen}}, \bibinfo {author} {\bibfnamefont
  {R.}~\bibnamefont {Golestanian}}, \bibinfo {author} {\bibfnamefont {U.~B.}\
  \bibnamefont {Kaupp}}, \bibinfo {author} {\bibfnamefont {L.}~\bibnamefont
  {Alvarez}}, \bibinfo {author} {\bibfnamefont {T.}~\bibnamefont {Ki{\o}rboe}},
  \bibinfo {author} {\bibfnamefont {E.}~\bibnamefont {Lauga}}, \bibinfo
  {author} {\bibfnamefont {W.~C.~K.}\ \bibnamefont {Poon}}, \bibinfo {author}
  {\bibfnamefont {A.}~\bibnamefont {DeSimone}}, \bibinfo {author}
  {\bibfnamefont {S.}~\bibnamefont {Mui{\~{n}}os-Landin}}, \bibinfo {author}
  {\bibfnamefont {A.}~\bibnamefont {Fischer}}, \bibinfo {author} {\bibfnamefont
  {N.~A.}\ \bibnamefont {Söker}}, \bibinfo {author} {\bibfnamefont
  {F.}~\bibnamefont {Cichos}}, \bibinfo {author} {\bibfnamefont
  {R.}~\bibnamefont {Kapral}}, \bibinfo {author} {\bibfnamefont
  {P.}~\bibnamefont {Gaspard}}, \bibinfo {author} {\bibfnamefont
  {M.}~\bibnamefont {Ripoll}}, \bibinfo {author} {\bibfnamefont
  {F.}~\bibnamefont {Sagues}}, \bibinfo {author} {\bibfnamefont
  {A.}~\bibnamefont {Doostmohammadi}}, \bibinfo {author} {\bibfnamefont
  {J.~M.}\ \bibnamefont {Yeomans}}, \bibinfo {author} {\bibfnamefont {I.~S.}\
  \bibnamefont {Aranson}}, \bibinfo {author} {\bibfnamefont {C.}~\bibnamefont
  {Bechinger}}, \bibinfo {author} {\bibfnamefont {H.}~\bibnamefont {Stark}},
  \bibinfo {author} {\bibfnamefont {C.~K.}\ \bibnamefont {Hemelrijk}}, \bibinfo
  {author} {\bibfnamefont {F.~J.}\ \bibnamefont {Nedelec}}, \bibinfo {author}
  {\bibfnamefont {T.}~\bibnamefont {Sarkar}}, \bibinfo {author} {\bibfnamefont
  {T.}~\bibnamefont {Aryaksama}}, \bibinfo {author} {\bibfnamefont
  {M.}~\bibnamefont {Lacroix}}, \bibinfo {author} {\bibfnamefont
  {G.}~\bibnamefont {Duclos}}, \bibinfo {author} {\bibfnamefont
  {V.}~\bibnamefont {Yashunsky}}, \bibinfo {author} {\bibfnamefont
  {P.}~\bibnamefont {Silberzan}}, \bibinfo {author} {\bibfnamefont
  {M.}~\bibnamefont {Arroyo}},\ and\ \bibinfo {author} {\bibfnamefont
  {S.}~\bibnamefont {Kale}},\ }\bibfield  {title} {\bibinfo {title} {The 2020
  motile active matter roadmap},\ }\href
  {https://doi.org/10.1088/1361-648x/ab6348} {\bibfield  {journal} {\bibinfo
  {journal} {Journal of Physics: Condensed Matter}\ }\textbf {\bibinfo {volume}
  {32}},\ \bibinfo {pages} {193001} (\bibinfo {year} {2020})}\BibitemShut
  {NoStop}%
\bibitem [{\citenamefont {Nelson}\ \emph {et~al.}(2010)\citenamefont {Nelson},
  \citenamefont {Kaliakastos},\ and\ \citenamefont {Abbott}}]{nelson_2010}%
  \BibitemOpen
  \bibfield  {author} {\bibinfo {author} {\bibfnamefont {B.~J.}\ \bibnamefont
  {Nelson}}, \bibinfo {author} {\bibfnamefont {I.~K.}\ \bibnamefont
  {Kaliakastos}},\ and\ \bibinfo {author} {\bibfnamefont {J.~J.}\ \bibnamefont
  {Abbott}},\ }\bibfield  {title} {\bibinfo {title} {Microrobots for minimally
  invasive medicine},\ }\href
  {https://doi.org/10.1146/annurev-bioeng-010510-103409} {\bibfield  {journal}
  {\bibinfo  {journal} {Ann. Rev. Biomed. Eng.}\ }\textbf {\bibinfo {volume}
  {12}},\ \bibinfo {pages} {55} (\bibinfo {year} {2010})}\BibitemShut {NoStop}%
\bibitem [{\citenamefont {Li}\ \emph {et~al.}(2017)\citenamefont {Li},
  \citenamefont {Esteban-Fern{\'a}ndez~de {\'A}vila}, \citenamefont {Gao},
  \citenamefont {Zhang},\ and\ \citenamefont {Wang}}]{li_2017}%
  \BibitemOpen
  \bibfield  {author} {\bibinfo {author} {\bibfnamefont {J.}~\bibnamefont
  {Li}}, \bibinfo {author} {\bibfnamefont {B.}~\bibnamefont
  {Esteban-Fern{\'a}ndez~de {\'A}vila}}, \bibinfo {author} {\bibfnamefont
  {W.}~\bibnamefont {Gao}}, \bibinfo {author} {\bibfnamefont {L.}~\bibnamefont
  {Zhang}},\ and\ \bibinfo {author} {\bibfnamefont {J.}~\bibnamefont {Wang}},\
  }\bibfield  {title} {\bibinfo {title} {Micro/nanorobots for biomedicine:
  Delivery, surgery, sensing, and detoxification},\ }\bibfield  {journal}
  {\bibinfo  {journal} {Science Robotics}\ }\textbf {\bibinfo {volume} {2}},\
  \href {https://doi.org/10.1126/scirobotics.aam6431}
  {10.1126/scirobotics.aam6431} (\bibinfo {year} {2017})\BibitemShut {NoStop}%
\bibitem [{\citenamefont {Hu}\ \emph {et~al.}(2018)\citenamefont {Hu},
  \citenamefont {Pan\'e},\ and\ \citenamefont {Nelson}}]{hu_2018}%
  \BibitemOpen
  \bibfield  {author} {\bibinfo {author} {\bibfnamefont {C.}~\bibnamefont
  {Hu}}, \bibinfo {author} {\bibfnamefont {S.}~\bibnamefont {Pan\'e}},\ and\
  \bibinfo {author} {\bibfnamefont {B.~J.}\ \bibnamefont {Nelson}},\ }\bibfield
   {title} {\bibinfo {title} {Soft micro- and nanorobotics},\ }\href
  {https://doi.org/10.1146/annurev-control-060117-104947} {\bibfield  {journal}
  {\bibinfo  {journal} {Annual Review of Control, Robotics, and Autonomous
  Systems}\ }\textbf {\bibinfo {volume} {1}},\ \bibinfo {pages} {53} (\bibinfo
  {year} {2018})}\BibitemShut {NoStop}%
\bibitem [{\citenamefont {Ghosh}\ \emph {et~al.}(2020)\citenamefont {Ghosh},
  \citenamefont {Xu}, \citenamefont {Gupta},\ and\ \citenamefont
  {Gracias}}]{ghosh_2020}%
  \BibitemOpen
  \bibfield  {author} {\bibinfo {author} {\bibfnamefont {A.}~\bibnamefont
  {Ghosh}}, \bibinfo {author} {\bibfnamefont {W.}~\bibnamefont {Xu}}, \bibinfo
  {author} {\bibfnamefont {N.}~\bibnamefont {Gupta}},\ and\ \bibinfo {author}
  {\bibfnamefont {D.~H.}\ \bibnamefont {Gracias}},\ }\bibfield  {title}
  {\bibinfo {title} {Active matter therapeutics},\ }\href
  {https://doi.org/10.1016/j.nantod.2019.100836} {\bibfield  {journal}
  {\bibinfo  {journal} {Nano Today}\ }\textbf {\bibinfo {volume} {31}},\
  \bibinfo {pages} {100836} (\bibinfo {year} {2020})}\BibitemShut {NoStop}%
\bibitem [{\citenamefont {Sundararajan}\ \emph {et~al.}(2008)\citenamefont
  {Sundararajan}, \citenamefont {Lammert}, \citenamefont {Zudans},
  \citenamefont {Crespi},\ and\ \citenamefont {Sen}}]{sundararajan_2008}%
  \BibitemOpen
  \bibfield  {author} {\bibinfo {author} {\bibfnamefont {S.}~\bibnamefont
  {Sundararajan}}, \bibinfo {author} {\bibfnamefont {P.~E.}\ \bibnamefont
  {Lammert}}, \bibinfo {author} {\bibfnamefont {A.~W.}\ \bibnamefont {Zudans}},
  \bibinfo {author} {\bibfnamefont {V.~H.}\ \bibnamefont {Crespi}},\ and\
  \bibinfo {author} {\bibfnamefont {A.}~\bibnamefont {Sen}},\ }\bibfield
  {title} {\bibinfo {title} {Catalytic motors for transport of colloidal
  cargo},\ }\href {https://doi.org/10.1021/nl072275j} {\bibfield  {journal}
  {\bibinfo  {journal} {Nano Lett.}\ }\textbf {\bibinfo {volume} {8}},\
  \bibinfo {pages} {1271} (\bibinfo {year} {2008})}\BibitemShut {NoStop}%
\bibitem [{\citenamefont {Burdick}\ \emph {et~al.}(2008)\citenamefont
  {Burdick}, \citenamefont {Laocharoenshuk}, \citenamefont {Wheat},
  \citenamefont {Posner},\ and\ \citenamefont {J.Wang}}]{burdick_2008}%
  \BibitemOpen
  \bibfield  {author} {\bibinfo {author} {\bibfnamefont {J.}~\bibnamefont
  {Burdick}}, \bibinfo {author} {\bibfnamefont {R.}~\bibnamefont
  {Laocharoenshuk}}, \bibinfo {author} {\bibfnamefont {P.~M.}\ \bibnamefont
  {Wheat}}, \bibinfo {author} {\bibfnamefont {J.~D.}\ \bibnamefont {Posner}},\
  and\ \bibinfo {author} {\bibnamefont {J.Wang}},\ }\bibfield  {title}
  {\bibinfo {title} {Synthetic nanomotors in microchannel networks: Directional
  microchip motion and controlled manipulation of cargo},\ }\href
  {https://doi.org/10.1021/ja803529u} {\bibfield  {journal} {\bibinfo
  {journal} {J. Am. Chem. Soc.}\ }\textbf {\bibinfo {volume} {130}},\ \bibinfo
  {pages} {8164} (\bibinfo {year} {2008})}\BibitemShut {NoStop}%
\bibitem [{\citenamefont {Fusco}\ \emph {et~al.}(2014)\citenamefont {Fusco},
  \citenamefont {Ullrich}, \citenamefont {Pokki}, \citenamefont
  {Chatzipirpiridis}, \citenamefont {{\"{O}}zkale}, \citenamefont {Sivaraman},
  \citenamefont {Ergeneman}, \citenamefont {Pan{\'{e}}},\ and\ \citenamefont
  {Nelson}}]{fusco_2014}%
  \BibitemOpen
  \bibfield  {author} {\bibinfo {author} {\bibfnamefont {S.}~\bibnamefont
  {Fusco}}, \bibinfo {author} {\bibfnamefont {F.}~\bibnamefont {Ullrich}},
  \bibinfo {author} {\bibfnamefont {J.}~\bibnamefont {Pokki}}, \bibinfo
  {author} {\bibfnamefont {G.}~\bibnamefont {Chatzipirpiridis}}, \bibinfo
  {author} {\bibfnamefont {B.}~\bibnamefont {{\"{O}}zkale}}, \bibinfo {author}
  {\bibfnamefont {K.~M.}\ \bibnamefont {Sivaraman}}, \bibinfo {author}
  {\bibfnamefont {O.}~\bibnamefont {Ergeneman}}, \bibinfo {author}
  {\bibfnamefont {S.}~\bibnamefont {Pan{\'{e}}}},\ and\ \bibinfo {author}
  {\bibfnamefont {B.~J.}\ \bibnamefont {Nelson}},\ }\bibfield  {title}
  {\bibinfo {title} {Microrobots: a new era in ocular drug delivery},\ }\href
  {https://doi.org/10.1517/17425247.2014.938633} {\bibfield  {journal}
  {\bibinfo  {journal} {Expert Opin Drug Deliv.}\ }\textbf {\bibinfo {volume}
  {11}},\ \bibinfo {pages} {1815} (\bibinfo {year} {2014})}\BibitemShut
  {NoStop}%
\bibitem [{\citenamefont {Luo}\ \emph {et~al.}(2018)\citenamefont {Luo},
  \citenamefont {Feng}, \citenamefont {Wang},\ and\ \citenamefont
  {Guan}}]{luo_2018}%
  \BibitemOpen
  \bibfield  {author} {\bibinfo {author} {\bibfnamefont {M.}~\bibnamefont
  {Luo}}, \bibinfo {author} {\bibfnamefont {Y.}~\bibnamefont {Feng}}, \bibinfo
  {author} {\bibfnamefont {T.}~\bibnamefont {Wang}},\ and\ \bibinfo {author}
  {\bibfnamefont {J.}~\bibnamefont {Guan}},\ }\bibfield  {title} {\bibinfo
  {title} {Micro-/nanorobots at work in active drug delivery},\ }\href
  {https://doi.org/10.1002/adfm.201706100} {\bibfield  {journal} {\bibinfo
  {journal} {Advanced Functional Materials}\ }\textbf {\bibinfo {volume}
  {28}},\ \bibinfo {pages} {1706100} (\bibinfo {year} {2018})}\BibitemShut
  {NoStop}%
\bibitem [{\citenamefont {Sonntag}\ \emph {et~al.}(2019)\citenamefont
  {Sonntag}, \citenamefont {Simmchen},\ and\ \citenamefont
  {Magdanz}}]{sonntag_2019}%
  \BibitemOpen
  \bibfield  {author} {\bibinfo {author} {\bibfnamefont {L.}~\bibnamefont
  {Sonntag}}, \bibinfo {author} {\bibfnamefont {J.}~\bibnamefont {Simmchen}},\
  and\ \bibinfo {author} {\bibfnamefont {V.}~\bibnamefont {Magdanz}},\
  }\bibfield  {title} {\bibinfo {title} {Nano-and micromotors designed for
  cancer therapy},\ }\href {https://doi.org/10.3390/molecules24183410}
  {\bibfield  {journal} {\bibinfo  {journal} {Molecules}\ }\textbf {\bibinfo
  {volume} {24}},\ \bibinfo {pages} {3410} (\bibinfo {year}
  {2019})}\BibitemShut {NoStop}%
\bibitem [{\citenamefont {Wu}\ \emph {et~al.}(2010)\citenamefont {Wu},
  \citenamefont {Balasubramanian}, \citenamefont {Kagan}, \citenamefont
  {Manesh}, \citenamefont {Campuzano},\ and\ \citenamefont
  {Wang}}]{balasubramanian_2010}%
  \BibitemOpen
  \bibfield  {author} {\bibinfo {author} {\bibfnamefont {J.}~\bibnamefont
  {Wu}}, \bibinfo {author} {\bibfnamefont {S.}~\bibnamefont {Balasubramanian}},
  \bibinfo {author} {\bibfnamefont {D.}~\bibnamefont {Kagan}}, \bibinfo
  {author} {\bibfnamefont {K.~M.}\ \bibnamefont {Manesh}}, \bibinfo {author}
  {\bibfnamefont {S.}~\bibnamefont {Campuzano}},\ and\ \bibinfo {author}
  {\bibfnamefont {J.}~\bibnamefont {Wang}},\ }\bibfield  {title} {\bibinfo
  {title} {Motion--based dna detection using catalytic nanomotors},\ }\href
  {https://doi.org/10.1038/ncomms1035} {\bibfield  {journal} {\bibinfo
  {journal} {Nat. Commun.}\ }\textbf {\bibinfo {volume} {1}},\ \bibinfo {pages}
  {36} (\bibinfo {year} {2010})}\BibitemShut {NoStop}%
\bibitem [{\citenamefont {Campuzano}\ \emph {et~al.}(2011)\citenamefont
  {Campuzano}, \citenamefont {Kagan}, \citenamefont {Orozco},\ and\
  \citenamefont {J.Wang}}]{campuzano_2011}%
  \BibitemOpen
  \bibfield  {author} {\bibinfo {author} {\bibfnamefont {S.}~\bibnamefont
  {Campuzano}}, \bibinfo {author} {\bibfnamefont {D.}~\bibnamefont {Kagan}},
  \bibinfo {author} {\bibfnamefont {J.}~\bibnamefont {Orozco}},\ and\ \bibinfo
  {author} {\bibnamefont {J.Wang}},\ }\bibfield  {title} {\bibinfo {title}
  {Motion-driven sensing and biosensing using electrochemically propelled
  nanomotors},\ }\href {https://doi.org/10.1039/C1AN15599G} {\bibfield
  {journal} {\bibinfo  {journal} {Analyst}\ }\textbf {\bibinfo {volume}
  {136}},\ \bibinfo {pages} {4621} (\bibinfo {year} {2011})}\BibitemShut
  {NoStop}%
\bibitem [{\citenamefont {Soler}\ \emph {et~al.}(2013)\citenamefont {Soler},
  \citenamefont {Magdanz}, \citenamefont {Fomin}, \citenamefont {Sanchez},\
  and\ \citenamefont {Schmidt}}]{soler_2013}%
  \BibitemOpen
  \bibfield  {author} {\bibinfo {author} {\bibfnamefont {L.}~\bibnamefont
  {Soler}}, \bibinfo {author} {\bibfnamefont {V.}~\bibnamefont {Magdanz}},
  \bibinfo {author} {\bibfnamefont {V.~M.}\ \bibnamefont {Fomin}}, \bibinfo
  {author} {\bibfnamefont {S.}~\bibnamefont {Sanchez}},\ and\ \bibinfo {author}
  {\bibfnamefont {O.~G.}\ \bibnamefont {Schmidt}},\ }\bibfield  {title}
  {\bibinfo {title} {Self-propelled micromotors for cleaning polluted water},\
  }\href {https://doi.org/10.1021/nn405075d} {\bibfield  {journal} {\bibinfo
  {journal} {ACS Nano}\ }\textbf {\bibinfo {volume} {7}},\ \bibinfo {pages}
  {9611} (\bibinfo {year} {2013})}\BibitemShut {NoStop}%
\bibitem [{\citenamefont {Nadal}\ and\ \citenamefont
  {Lauga}(2014)}]{nadal_2014}%
  \BibitemOpen
  \bibfield  {author} {\bibinfo {author} {\bibfnamefont {F.}~\bibnamefont
  {Nadal}}\ and\ \bibinfo {author} {\bibfnamefont {E.}~\bibnamefont {Lauga}},\
  }\bibfield  {title} {\bibinfo {title} {Asymmetric steady streaming as a
  mechanism for acoustic propulsion of rigid bodies},\ }\href
  {https://doi.org/10.1063/1.4891446} {\bibfield  {journal} {\bibinfo
  {journal} {Phys. Fluids}\ }\textbf {\bibinfo {volume} {26}},\ \bibinfo
  {pages} {082001} (\bibinfo {year} {2014})}\BibitemShut {NoStop}%
\bibitem [{\citenamefont {Najafi}\ and\ \citenamefont
  {Golestanian}(2004)}]{najafi_2004}%
  \BibitemOpen
  \bibfield  {author} {\bibinfo {author} {\bibfnamefont {A.}~\bibnamefont
  {Najafi}}\ and\ \bibinfo {author} {\bibfnamefont {R.}~\bibnamefont
  {Golestanian}},\ }\bibfield  {title} {\bibinfo {title} {Simple swimmer at low
  reynolds number: Three linked spheres},\ }\href
  {https://doi.org/10.1103/PhysRevE.69.062901} {\bibfield  {journal} {\bibinfo
  {journal} {Phys. Rev. E}\ }\textbf {\bibinfo {volume} {69}},\ \bibinfo
  {pages} {062901} (\bibinfo {year} {2004})}\BibitemShut {NoStop}%
\bibitem [{\citenamefont {Leoni}\ \emph {et~al.}(2009)\citenamefont {Leoni},
  \citenamefont {Kotar}, \citenamefont {Bassetti}, \citenamefont {Cicuta},\
  and\ \citenamefont {Lagomarsino}}]{leoni_2009}%
  \BibitemOpen
  \bibfield  {author} {\bibinfo {author} {\bibfnamefont {M.}~\bibnamefont
  {Leoni}}, \bibinfo {author} {\bibfnamefont {J.}~\bibnamefont {Kotar}},
  \bibinfo {author} {\bibfnamefont {B.}~\bibnamefont {Bassetti}}, \bibinfo
  {author} {\bibfnamefont {P.}~\bibnamefont {Cicuta}},\ and\ \bibinfo {author}
  {\bibfnamefont {M.~C.}\ \bibnamefont {Lagomarsino}},\ }\bibfield  {title}
  {\bibinfo {title} {A basic swimmer at low reynolds number},\ }\href
  {https://doi.org/10.1039/B812393D} {\bibfield  {journal} {\bibinfo  {journal}
  {Soft Matter}\ }\textbf {\bibinfo {volume} {5}},\ \bibinfo {pages} {472}
  (\bibinfo {year} {2009})}\BibitemShut {NoStop}%
\bibitem [{\citenamefont {Grosjean}\ \emph {et~al.}(2016)\citenamefont
  {Grosjean}, \citenamefont {Hubert}, \citenamefont {Lagubeau},\ and\
  \citenamefont {Vandewalle}}]{grosjean_2016}%
  \BibitemOpen
  \bibfield  {author} {\bibinfo {author} {\bibfnamefont {G.}~\bibnamefont
  {Grosjean}}, \bibinfo {author} {\bibfnamefont {M.}~\bibnamefont {Hubert}},
  \bibinfo {author} {\bibfnamefont {G.}~\bibnamefont {Lagubeau}},\ and\
  \bibinfo {author} {\bibfnamefont {N.}~\bibnamefont {Vandewalle}},\ }\bibfield
   {title} {\bibinfo {title} {Realization of the najafi-golestanian
  microswimmer},\ }\href {https://doi.org/10.1103/PhysRevE.94.021101}
  {\bibfield  {journal} {\bibinfo  {journal} {Phys. Rev. E}\ }\textbf {\bibinfo
  {volume} {94}},\ \bibinfo {pages} {021101} (\bibinfo {year}
  {2016})}\BibitemShut {NoStop}%
\bibitem [{\citenamefont {Box}\ \emph {et~al.}(2017)\citenamefont {Box},
  \citenamefont {Han}, \citenamefont {Tipton},\ and\ \citenamefont
  {Mullin}}]{box_2017}%
  \BibitemOpen
  \bibfield  {author} {\bibinfo {author} {\bibfnamefont {F.}~\bibnamefont
  {Box}}, \bibinfo {author} {\bibfnamefont {E.}~\bibnamefont {Han}}, \bibinfo
  {author} {\bibfnamefont {C.~R.}\ \bibnamefont {Tipton}},\ and\ \bibinfo
  {author} {\bibfnamefont {T.}~\bibnamefont {Mullin}},\ }\bibfield  {title}
  {\bibinfo {title} {On the motion of linked spheres in a stokes flow},\ }\href
  {https://doi.org/10.1007/s00348-017-2321-2} {\bibfield  {journal} {\bibinfo
  {journal} {Experiments in Fluids}\ }\textbf {\bibinfo {volume} {58}},\
  \bibinfo {pages} {29} (\bibinfo {year} {2017})}\BibitemShut {NoStop}%
\bibitem [{\citenamefont {Elder}\ \emph {et~al.}(2021)\citenamefont {Elder},
  \citenamefont {Zou}, \citenamefont {Ghosh}, \citenamefont {Silverberg},
  \citenamefont {Greenwood}, \citenamefont {Demir}, \citenamefont {Su},
  \citenamefont {Pak},\ and\ \citenamefont {Kong}}]{Elder_2021}%
  \BibitemOpen
  \bibfield  {author} {\bibinfo {author} {\bibfnamefont {B.}~\bibnamefont
  {Elder}}, \bibinfo {author} {\bibfnamefont {Z.}~\bibnamefont {Zou}}, \bibinfo
  {author} {\bibfnamefont {S.}~\bibnamefont {Ghosh}}, \bibinfo {author}
  {\bibfnamefont {O.}~\bibnamefont {Silverberg}}, \bibinfo {author}
  {\bibfnamefont {T.~E.}\ \bibnamefont {Greenwood}}, \bibinfo {author}
  {\bibfnamefont {E.}~\bibnamefont {Demir}}, \bibinfo {author} {\bibfnamefont
  {V.~S.-E.}\ \bibnamefont {Su}}, \bibinfo {author} {\bibfnamefont {O.~S.}\
  \bibnamefont {Pak}},\ and\ \bibinfo {author} {\bibfnamefont {Y.~L.}\
  \bibnamefont {Kong}},\ }\bibfield  {title} {\bibinfo {title} {A 3d-printed
  self-learning three-linked-sphere robot for autonomous confined-space
  navigation},\ }\href {https://doi.org/https://doi.org/10.1002/aisy.202100039}
  {\bibfield  {journal} {\bibinfo  {journal} {Advanced Intelligent Systems}\
  }\textbf {\bibinfo {volume} {3}},\ \bibinfo {pages} {2100039} (\bibinfo
  {year} {2021})}\BibitemShut {NoStop}%
\bibitem [{\citenamefont {Felderhof}(2006)}]{felderhof_2006}%
  \BibitemOpen
  \bibfield  {author} {\bibinfo {author} {\bibfnamefont {B.~U.}\ \bibnamefont
  {Felderhof}},\ }\bibfield  {title} {\bibinfo {title} {The swimming of
  animalcules},\ }\href {https://doi.org/10.1063/1.2204633} {\bibfield
  {journal} {\bibinfo  {journal} {Physics of Fluids}\ }\textbf {\bibinfo
  {volume} {18}},\ \bibinfo {pages} {063101} (\bibinfo {year}
  {2006})}\BibitemShut {NoStop}%
\bibitem [{\citenamefont {Pooley}\ \emph {et~al.}(2007)\citenamefont {Pooley},
  \citenamefont {Alexander},\ and\ \citenamefont {Yeomans}}]{pooley_2007}%
  \BibitemOpen
  \bibfield  {author} {\bibinfo {author} {\bibfnamefont {C.~M.}\ \bibnamefont
  {Pooley}}, \bibinfo {author} {\bibfnamefont {G.~P.}\ \bibnamefont
  {Alexander}},\ and\ \bibinfo {author} {\bibfnamefont {J.~M.}\ \bibnamefont
  {Yeomans}},\ }\href@noop {} {\bibfield  {journal} {\bibinfo  {journal} {Phys.
  Rev. Lett.}\ }\textbf {\bibinfo {volume} {99}},\ \bibinfo {pages} {228103}
  (\bibinfo {year} {2007})}\BibitemShut {NoStop}%
\bibitem [{\citenamefont {Golestanian}\ and\ \citenamefont
  {Ajdari}(2008)}]{golestanian_2008}%
  \BibitemOpen
  \bibfield  {author} {\bibinfo {author} {\bibfnamefont {R.}~\bibnamefont
  {Golestanian}}\ and\ \bibinfo {author} {\bibfnamefont {A.}~\bibnamefont
  {Ajdari}},\ }\bibfield  {title} {\bibinfo {title} {Analytic results for the
  three-sphere swimmer at low reynolds number},\ }\href
  {https://doi.org/10.1103/PhysRevE.77.036308} {\bibfield  {journal} {\bibinfo
  {journal} {Phys. Rev. E}\ }\textbf {\bibinfo {volume} {77}},\ \bibinfo
  {pages} {036308} (\bibinfo {year} {2008})}\BibitemShut {NoStop}%
\bibitem [{\citenamefont {Alouges}\ \emph {et~al.}(2008)\citenamefont
  {Alouges}, \citenamefont {DeSimone},\ and\ \citenamefont
  {Lefebvre}}]{alouges_2008}%
  \BibitemOpen
  \bibfield  {author} {\bibinfo {author} {\bibfnamefont {F.}~\bibnamefont
  {Alouges}}, \bibinfo {author} {\bibfnamefont {A.}~\bibnamefont {DeSimone}},\
  and\ \bibinfo {author} {\bibfnamefont {A.}~\bibnamefont {Lefebvre}},\
  }\bibfield  {title} {\bibinfo {title} {Optimal strokes for low reynolds
  number swimmers: An example},\ }\href
  {https://doi.org/10.1007/s00332-007-9013-7} {\bibfield  {journal} {\bibinfo
  {journal} {J Nonlinear Sci}\ }\textbf {\bibinfo {volume} {18}},\ \bibinfo
  {pages} {277} (\bibinfo {year} {2008})}\BibitemShut {NoStop}%
\bibitem [{\citenamefont {Alexander}\ \emph {et~al.}(2008)\citenamefont
  {Alexander}, \citenamefont {Pooley},\ and\ \citenamefont
  {Yeomans}}]{alexander_2008}%
  \BibitemOpen
  \bibfield  {author} {\bibinfo {author} {\bibfnamefont {G.~P.}\ \bibnamefont
  {Alexander}}, \bibinfo {author} {\bibfnamefont {C.~M.}\ \bibnamefont
  {Pooley}},\ and\ \bibinfo {author} {\bibfnamefont {J.~M.}\ \bibnamefont
  {Yeomans}},\ }\bibfield  {title} {\bibinfo {title} {Scattering of low
  reynolds number swimmers},\ }\href
  {https://doi.org/10.1103/PhysRevE.78.045302} {\bibfield  {journal} {\bibinfo
  {journal} {Phys. Rev. E}\ }\textbf {\bibinfo {volume} {78}},\ \bibinfo
  {pages} {045302} (\bibinfo {year} {2008})}\BibitemShut {NoStop}%
\bibitem [{\citenamefont {Alexander}\ \emph {et~al.}(2009)\citenamefont
  {Alexander}, \citenamefont {Pooley},\ and\ \citenamefont
  {Yeomans}}]{alexander_2009}%
  \BibitemOpen
  \bibfield  {author} {\bibinfo {author} {\bibfnamefont {G.~P.}\ \bibnamefont
  {Alexander}}, \bibinfo {author} {\bibfnamefont {C.~M.}\ \bibnamefont
  {Pooley}},\ and\ \bibinfo {author} {\bibfnamefont {J.~M.}\ \bibnamefont
  {Yeomans}},\ }\bibfield  {title} {\bibinfo {title} {Hydrodynamics of linked
  sphere model swimmers},\ }\href
  {https://doi.org/10.1088/0953-8984/21/20/204108} {\bibfield  {journal}
  {\bibinfo  {journal} {Journal of Physics: Condensed Matter}\ }\textbf
  {\bibinfo {volume} {21}},\ \bibinfo {pages} {204108} (\bibinfo {year}
  {2009})}\BibitemShut {NoStop}%
\bibitem [{\citenamefont {Alouges}\ \emph {et~al.}(2009)\citenamefont
  {Alouges}, \citenamefont {DeSimone},\ and\ \citenamefont
  {Lefebvre}}]{alouges_2009}%
  \BibitemOpen
  \bibfield  {author} {\bibinfo {author} {\bibfnamefont {F.}~\bibnamefont
  {Alouges}}, \bibinfo {author} {\bibfnamefont {A.}~\bibnamefont {DeSimone}},\
  and\ \bibinfo {author} {\bibfnamefont {A.}~\bibnamefont {Lefebvre}},\
  }\bibfield  {title} {\bibinfo {title} {Optimal strokes for axisymmetric
  microswimmers},\ }\href {https://doi.org/10.1140/epje/i2008-10406-4}
  {\bibfield  {journal} {\bibinfo  {journal} {Eur. Phys. J. E}\ }\textbf
  {\bibinfo {volume} {28}},\ \bibinfo {pages} {279} (\bibinfo {year}
  {2009})}\BibitemShut {NoStop}%
\bibitem [{\citenamefont {Vladimirov}(2013)}]{vladimirov_2013}%
  \BibitemOpen
  \bibfield  {author} {\bibinfo {author} {\bibfnamefont {V.~A.}\ \bibnamefont
  {Vladimirov}},\ }\bibfield  {title} {\bibinfo {title} {On the self-propulsion
  of an $n$ -sphere micro-robot},\ }\href
  {https://doi.org/10.1017/jfm.2012.501} {\bibfield  {journal} {\bibinfo
  {journal} {Journal of Fluid Mechanics}\ }\textbf {\bibinfo {volume} {716}},\
  \bibinfo {pages} {R1} (\bibinfo {year} {2013})}\BibitemShut {NoStop}%
\bibitem [{\citenamefont {Ferlderhof}(2014)}]{felderhof_2014}%
  \BibitemOpen
  \bibfield  {author} {\bibinfo {author} {\bibfnamefont {U.}~\bibnamefont
  {Ferlderhof}},\ }\bibfield  {title} {\bibinfo {title} {Swimming of an
  assembly of rigid spheres at low reynolds number},\ }\href@noop {} {\bibfield
   {journal} {\bibinfo  {journal} {Eur. Phys. J. E}\ }\textbf {\bibinfo
  {volume} {37}},\ \bibinfo {pages} {110} (\bibinfo {year} {2014})}\BibitemShut
  {NoStop}%
\bibitem [{\citenamefont {Montino}\ and\ \citenamefont
  {DeSimone}(2015)}]{montino_2015}%
  \BibitemOpen
  \bibfield  {author} {\bibinfo {author} {\bibfnamefont {A.}~\bibnamefont
  {Montino}}\ and\ \bibinfo {author} {\bibfnamefont {A.}~\bibnamefont
  {DeSimone}},\ }\bibfield  {title} {\bibinfo {title} {Three-sphere
  low-reynolds-number swimmer with a passive elastic arm},\ }\href
  {https://doi.org/10.1140/epje/i2015-15042-3} {\bibfield  {journal} {\bibinfo
  {journal} {The European Physical Journal E}\ }\textbf {\bibinfo {volume}
  {38}},\ \bibinfo {pages} {42} (\bibinfo {year} {2015})}\BibitemShut {NoStop}%
\bibitem [{\citenamefont {Montino}\ and\ \citenamefont
  {DeSimone}(2017)}]{montino_2017}%
  \BibitemOpen
  \bibfield  {author} {\bibinfo {author} {\bibfnamefont {A.}~\bibnamefont
  {Montino}}\ and\ \bibinfo {author} {\bibfnamefont {A.}~\bibnamefont
  {DeSimone}},\ }\bibfield  {title} {\bibinfo {title} {Dynamics and optimal
  actuation of a three-sphere low-reynolds-number swimmer with muscle-like
  arms},\ }\href {https://doi.org/10.1007/s10440-016-0087-9} {\bibfield
  {journal} {\bibinfo  {journal} {Acta Applicandae Mathematicae}\ }\textbf
  {\bibinfo {volume} {149}},\ \bibinfo {pages} {53} (\bibinfo {year}
  {2017})}\BibitemShut {NoStop}%
\bibitem [{\citenamefont {Wang}(2019)}]{wang_2019}%
  \BibitemOpen
  \bibfield  {author} {\bibinfo {author} {\bibfnamefont {Q.}~\bibnamefont
  {Wang}},\ }\bibfield  {title} {\bibinfo {title} {Optimal strokes of low
  reynolds number linked-sphere swimmers},\ }\href
  {https://doi.org/10.3390/app9194023} {\bibfield  {journal} {\bibinfo
  {journal} {Applied Sciences}\ }\textbf {\bibinfo {volume} {9}},\ \bibinfo
  {pages} {4023} (\bibinfo {year} {2019})}\BibitemShut {NoStop}%
\bibitem [{\citenamefont {Earl}\ \emph {et~al.}(2007)\citenamefont {Earl},
  \citenamefont {Pooley}, \citenamefont {Ryder}, \citenamefont {Bredberg},\
  and\ \citenamefont {Yeomans}}]{earl_2007}%
  \BibitemOpen
  \bibfield  {author} {\bibinfo {author} {\bibfnamefont {D.~J.}\ \bibnamefont
  {Earl}}, \bibinfo {author} {\bibfnamefont {C.~M.}\ \bibnamefont {Pooley}},
  \bibinfo {author} {\bibfnamefont {J.~F.}\ \bibnamefont {Ryder}}, \bibinfo
  {author} {\bibfnamefont {I.}~\bibnamefont {Bredberg}},\ and\ \bibinfo
  {author} {\bibfnamefont {J.~M.}\ \bibnamefont {Yeomans}},\ }\bibfield
  {title} {\bibinfo {title} {Modeling microscopic swimmers at low reynolds
  number},\ }\href {https://doi.org/10.1063/1.2434160} {\bibfield  {journal}
  {\bibinfo  {journal} {J. Chem. Phys.}\ }\textbf {\bibinfo {volume} {126}},\
  \bibinfo {pages} {064703} (\bibinfo {year} {2007})}\BibitemShut {NoStop}%
\bibitem [{\citenamefont {Nasouri}\ \emph {et~al.}(2019)\citenamefont
  {Nasouri}, \citenamefont {Vilfan},\ and\ \citenamefont
  {Golestanian}}]{nasouri_2019}%
  \BibitemOpen
  \bibfield  {author} {\bibinfo {author} {\bibfnamefont {B.}~\bibnamefont
  {Nasouri}}, \bibinfo {author} {\bibfnamefont {A.}~\bibnamefont {Vilfan}},\
  and\ \bibinfo {author} {\bibfnamefont {R.}~\bibnamefont {Golestanian}},\
  }\bibfield  {title} {\bibinfo {title} {Efficiency limits of the three-sphere
  swimmer},\ }\href {https://doi.org/10.1103/PhysRevFluids.4.073101} {\bibfield
   {journal} {\bibinfo  {journal} {Phys. Rev. Fluids}\ }\textbf {\bibinfo
  {volume} {4}},\ \bibinfo {pages} {073101} (\bibinfo {year}
  {2019})}\BibitemShut {NoStop}%
\bibitem [{\citenamefont {Pickl}\ \emph {et~al.}(2012)\citenamefont {Pickl},
  \citenamefont {G{\"{o}}tz}, \citenamefont {Iglberger}, \citenamefont {Pande},
  \citenamefont {Mecke}, \citenamefont {Smith},\ and\ \citenamefont
  {R{\"{u}}de}}]{pickl_2012}%
  \BibitemOpen
  \bibfield  {author} {\bibinfo {author} {\bibfnamefont {K.}~\bibnamefont
  {Pickl}}, \bibinfo {author} {\bibfnamefont {J.}~\bibnamefont {G{\"{o}}tz}},
  \bibinfo {author} {\bibfnamefont {K.}~\bibnamefont {Iglberger}}, \bibinfo
  {author} {\bibfnamefont {J.}~\bibnamefont {Pande}}, \bibinfo {author}
  {\bibfnamefont {K.}~\bibnamefont {Mecke}}, \bibinfo {author} {\bibfnamefont
  {A.-S.}\ \bibnamefont {Smith}},\ and\ \bibinfo {author} {\bibfnamefont
  {U.}~\bibnamefont {R{\"{u}}de}},\ }\bibfield  {title} {\bibinfo {title} {All
  good things come in threes—three beads learn to swim with lattice boltzmann
  and a rigid body solver},\ }\href
  {https://doi.org/10.1016/j.jocs.2012.04.009} {\bibfield  {journal} {\bibinfo
  {journal} {Journal of Computational Science}\ }\textbf {\bibinfo {volume}
  {3}},\ \bibinfo {pages} {374} (\bibinfo {year} {2012})}\BibitemShut {NoStop}%
\bibitem [{\citenamefont {Pickl}\ \emph {et~al.}(2017)\citenamefont {Pickl},
  \citenamefont {Pande}, \citenamefont {K{\"{o}}stler}, \citenamefont
  {R{\"{u}}de},\ and\ \citenamefont {Smith}}]{pickl_2017}%
  \BibitemOpen
  \bibfield  {author} {\bibinfo {author} {\bibfnamefont {K.}~\bibnamefont
  {Pickl}}, \bibinfo {author} {\bibfnamefont {J.}~\bibnamefont {Pande}},
  \bibinfo {author} {\bibfnamefont {H.}~\bibnamefont {K{\"{o}}stler}}, \bibinfo
  {author} {\bibfnamefont {U.}~\bibnamefont {R{\"{u}}de}},\ and\ \bibinfo
  {author} {\bibfnamefont {A.-S.}\ \bibnamefont {Smith}},\ }\bibfield  {title}
  {\bibinfo {title} {Lattice boltzmann simulations of the bead-spring
  microswimmer with a responsive stroke{\textemdash}from an individual to
  swarms},\ }\href {https://doi.org/10.1088/1361-648x/aa5a40} {\bibfield
  {journal} {\bibinfo  {journal} {Journal of Physics: Condensed Matter}\
  }\textbf {\bibinfo {volume} {29}},\ \bibinfo {pages} {124001} (\bibinfo
  {year} {2017})}\BibitemShut {NoStop}%
\bibitem [{\citenamefont {Nunes~Lengler}(2021)}]{nunes_2021}%
  \BibitemOpen
  \bibfield  {author} {\bibinfo {author} {\bibfnamefont {H.}~\bibnamefont
  {Nunes~Lengler}},\ }\bibfield  {title} {\bibinfo {title} {The regularized
  stokeslets method applied to the three-sphere swimmer model},\ }\href
  {https://doi.org/10.1063/5.0040052} {\bibfield  {journal} {\bibinfo
  {journal} {Physics of Fluids}\ }\textbf {\bibinfo {volume} {33}},\ \bibinfo
  {pages} {032007} (\bibinfo {year} {2021})}\BibitemShut {NoStop}%
\bibitem [{\citenamefont {Brady}\ and\ \citenamefont
  {Bossis}(1988)}]{brady_1988}%
  \BibitemOpen
  \bibfield  {author} {\bibinfo {author} {\bibfnamefont {J.~F.}\ \bibnamefont
  {Brady}}\ and\ \bibinfo {author} {\bibfnamefont {G.}~\bibnamefont {Bossis}},\
  }\bibfield  {title} {\bibinfo {title} {Stokesian dynamics},\ }\href
  {https://doi.org/10.1146/annurev.fl.20.010188.000551} {\bibfield  {journal}
  {\bibinfo  {journal} {Annu. Rev. Fluid Mech.}\ }\textbf {\bibinfo {volume}
  {20}},\ \bibinfo {pages} {111} (\bibinfo {year} {1988})}\BibitemShut
  {NoStop}%
\bibitem [{\citenamefont {Swan}\ \emph {et~al.}(2011)\citenamefont {Swan},
  \citenamefont {Brady}, \citenamefont {Moore},\ and\ \citenamefont
  {Che174}}]{swan_2011}%
  \BibitemOpen
  \bibfield  {author} {\bibinfo {author} {\bibfnamefont {J.~W.}\ \bibnamefont
  {Swan}}, \bibinfo {author} {\bibfnamefont {J.~F.}\ \bibnamefont {Brady}},
  \bibinfo {author} {\bibfnamefont {R.~S.}\ \bibnamefont {Moore}},\ and\
  \bibinfo {author} {\bibnamefont {Che174}},\ }\bibfield  {title} {\bibinfo
  {title} {Modeling hydrodynamic self-propulsion with stokesian dynamics. or
  teaching stokesian dynamics to swim},\ }\href
  {https://doi.org/10.1063/1.3594790} {\bibfield  {journal} {\bibinfo
  {journal} {Phys. Fluids}\ }\textbf {\bibinfo {volume} {23}},\ \bibinfo
  {pages} {071901} (\bibinfo {year} {2011})}\BibitemShut {NoStop}%
\bibitem [{\citenamefont {Phillips}\ \emph {et~al.}(1988)\citenamefont
  {Phillips}, \citenamefont {Brady},\ and\ \citenamefont
  {Bossis}}]{bossis_1988}%
  \BibitemOpen
  \bibfield  {author} {\bibinfo {author} {\bibfnamefont {R.~J.}\ \bibnamefont
  {Phillips}}, \bibinfo {author} {\bibfnamefont {J.~F.}\ \bibnamefont
  {Brady}},\ and\ \bibinfo {author} {\bibfnamefont {G.}~\bibnamefont
  {Bossis}},\ }\bibfield  {title} {\bibinfo {title} {Hydrodynamic transport
  properties of hard--sphere dispersions. i. suspensions of freely mobile
  particles},\ }\href {https://doi.org/10.1063/1.866914} {\bibfield  {journal}
  {\bibinfo  {journal} {Phys. Fluids}\ }\textbf {\bibinfo {volume} {31}},\
  \bibinfo {pages} {3462} (\bibinfo {year} {1988})}\BibitemShut {NoStop}%
\bibitem [{\citenamefont {Foss}\ and\ \citenamefont {Brady}(2000)}]{foss_2000}%
  \BibitemOpen
  \bibfield  {author} {\bibinfo {author} {\bibfnamefont {D.~R.}\ \bibnamefont
  {Foss}}\ and\ \bibinfo {author} {\bibfnamefont {J.~F.}\ \bibnamefont
  {Brady}},\ }\bibfield  {title} {\bibinfo {title} {Structure, diffusion and
  rheology of brownian suspensions by stokesian dynamics simulation},\ }\href
  {https://doi.org/10.1017/S0022112099007557} {\bibfield  {journal} {\bibinfo
  {journal} {J. Fluid Mech.}\ }\textbf {\bibinfo {volume} {407}},\ \bibinfo
  {pages} {167} (\bibinfo {year} {2000})}\BibitemShut {NoStop}%
\bibitem [{\citenamefont {Phung}\ \emph {et~al.}(1996)\citenamefont {Phung},
  \citenamefont {Brady},\ and\ \citenamefont {Bossis}}]{phung_1996}%
  \BibitemOpen
  \bibfield  {author} {\bibinfo {author} {\bibfnamefont {T.~N.}\ \bibnamefont
  {Phung}}, \bibinfo {author} {\bibfnamefont {J.~F.}\ \bibnamefont {Brady}},\
  and\ \bibinfo {author} {\bibfnamefont {G.}~\bibnamefont {Bossis}},\
  }\bibfield  {title} {\bibinfo {title} {Stokesian dynamics simulation of
  brownian suspensions},\ }\href {https://doi.org/10.1017/S0022112096002170}
  {\bibfield  {journal} {\bibinfo  {journal} {J. Fluid Mech.}\ }\textbf
  {\bibinfo {volume} {313}},\ \bibinfo {pages} {181} (\bibinfo {year}
  {1996})}\BibitemShut {NoStop}%
\bibitem [{\citenamefont {Holmqvist}\ \emph {et~al.}(2010)\citenamefont
  {Holmqvist}, \citenamefont {Banchio},\ and\ \citenamefont
  {N{\"{a}}gele}}]{heinen_2010}%
  \BibitemOpen
  \bibfield  {author} {\bibinfo {author} {\bibfnamefont {M.~H.~P.}\
  \bibnamefont {Holmqvist}}, \bibinfo {author} {\bibfnamefont {A.~J.}\
  \bibnamefont {Banchio}},\ and\ \bibinfo {author} {\bibfnamefont
  {G.}~\bibnamefont {N{\"{a}}gele}},\ }\bibfield  {title} {\bibinfo {title}
  {Short-time diffusion of charge-stabilized colloidal particles: generic
  features},\ }\href {https://doi.org/10.1107/S002188981002724X} {\bibfield
  {journal} {\bibinfo  {journal} {J. Appl. Cryst.}\ }\textbf {\bibinfo {volume}
  {43}},\ \bibinfo {pages} {970} (\bibinfo {year} {2010})}\BibitemShut
  {NoStop}%
\bibitem [{\citenamefont {Heinen}\ \emph {et~al.}(2011)\citenamefont {Heinen},
  \citenamefont {Banchio},\ and\ \citenamefont {N{\"{a}}gele}}]{heinen_2011}%
  \BibitemOpen
  \bibfield  {author} {\bibinfo {author} {\bibfnamefont {M.}~\bibnamefont
  {Heinen}}, \bibinfo {author} {\bibfnamefont {A.~J.}\ \bibnamefont
  {Banchio}},\ and\ \bibinfo {author} {\bibfnamefont {G.}~\bibnamefont
  {N{\"{a}}gele}},\ }\bibfield  {title} {\bibinfo {title} {Short-time rheology
  and diffusion in suspensions of yukawa-type colloidal particles},\ }\href
  {https://doi.org/10.1063/1.3646962} {\bibfield  {journal} {\bibinfo
  {journal} {J. Chem. Phys.}\ }\textbf {\bibinfo {volume} {135}},\ \bibinfo
  {pages} {154504} (\bibinfo {year} {2011})}\BibitemShut {NoStop}%
\bibitem [{\citenamefont {Banchio}\ \emph {et~al.}(2018)\citenamefont
  {Banchio}, \citenamefont {Heinen}, \citenamefont {Holmqvist},\ and\
  \citenamefont {N{\"{a}}gele}}]{banchio_2018}%
  \BibitemOpen
  \bibfield  {author} {\bibinfo {author} {\bibfnamefont {A.~J.}\ \bibnamefont
  {Banchio}}, \bibinfo {author} {\bibfnamefont {M.}~\bibnamefont {Heinen}},
  \bibinfo {author} {\bibfnamefont {P.}~\bibnamefont {Holmqvist}},\ and\
  \bibinfo {author} {\bibfnamefont {G.}~\bibnamefont {N{\"{a}}gele}},\
  }\bibfield  {title} {\bibinfo {title} {Short- and long-time diffusion and
  dynamic scaling in suspensions of charged colloidal particle},\ }\href
  {https://doi.org/10.1063/1.5017969} {\bibfield  {journal} {\bibinfo
  {journal} {J. Chem. Phys.}\ }\textbf {\bibinfo {volume} {148}},\ \bibinfo
  {pages} {134902} (\bibinfo {year} {2018})}\BibitemShut {NoStop}%
\bibitem [{\citenamefont {Ladd}(1994{\natexlab{a}})}]{ladd_1994a}%
  \BibitemOpen
  \bibfield  {author} {\bibinfo {author} {\bibfnamefont {A.~J.~C.}\
  \bibnamefont {Ladd}},\ }\bibfield  {title} {\bibinfo {title} {Numerical
  simulations of particulate suspensions via a discretized boltzmann equation.
  part 1. theoretical foundation},\ }\href
  {https://doi.org/10.1017/S0022112094001771} {\bibfield  {journal} {\bibinfo
  {journal} {Journal of Fluid Mechanics}\ }\textbf {\bibinfo {volume} {271}},\
  \bibinfo {pages} {285–309} (\bibinfo {year}
  {1994}{\natexlab{a}})}\BibitemShut {NoStop}%
\bibitem [{\citenamefont {Ladd}(1994{\natexlab{b}})}]{ladd_1994b}%
  \BibitemOpen
  \bibfield  {author} {\bibinfo {author} {\bibfnamefont {A.~J.~C.}\
  \bibnamefont {Ladd}},\ }\bibfield  {title} {\bibinfo {title} {Numerical
  simulations of particulate suspensions via a discretized boltzmann equation.
  part 2. numerical results},\ }\href
  {https://doi.org/10.1017/S0022112094001783} {\bibfield  {journal} {\bibinfo
  {journal} {Journal of Fluid Mechanics}\ }\textbf {\bibinfo {volume} {271}},\
  \bibinfo {pages} {311–339} (\bibinfo {year}
  {1994}{\natexlab{b}})}\BibitemShut {NoStop}%
\bibitem [{\citenamefont {Malevanets}\ and\ \citenamefont
  {Kapral}(1999)}]{malevanets_1999}%
  \BibitemOpen
  \bibfield  {author} {\bibinfo {author} {\bibfnamefont {A.}~\bibnamefont
  {Malevanets}}\ and\ \bibinfo {author} {\bibfnamefont {R.}~\bibnamefont
  {Kapral}},\ }\bibfield  {title} {\bibinfo {title} {Mesoscopic model for
  solvent dynamics},\ }\href {https://doi.org/10.1063/1.478857} {\bibfield
  {journal} {\bibinfo  {journal} {The Journal of Chemical Physics}\ }\textbf
  {\bibinfo {volume} {110}},\ \bibinfo {pages} {8605} (\bibinfo {year}
  {1999})}\BibitemShut {NoStop}%
\bibitem [{\citenamefont {Nguyen}\ and\ \citenamefont
  {Ladd}(2002)}]{nguyen_2002}%
  \BibitemOpen
  \bibfield  {author} {\bibinfo {author} {\bibfnamefont {N.-Q.}\ \bibnamefont
  {Nguyen}}\ and\ \bibinfo {author} {\bibfnamefont {A.~J.~C.}\ \bibnamefont
  {Ladd}},\ }\bibfield  {title} {\bibinfo {title} {Lubrication corrections for
  lattice-boltzmann simulations of particle suspensions},\ }\href
  {https://doi.org/10.1103/PhysRevE.66.046708} {\bibfield  {journal} {\bibinfo
  {journal} {Phys. Rev. E}\ }\textbf {\bibinfo {volume} {66}},\ \bibinfo
  {pages} {046708} (\bibinfo {year} {2002})}\BibitemShut {NoStop}%
\bibitem [{\citenamefont {Lighthill}(1976)}]{lighthill_1976}%
  \BibitemOpen
  \bibfield  {author} {\bibinfo {author} {\bibfnamefont {J.}~\bibnamefont
  {Lighthill}},\ }\bibfield  {title} {\bibinfo {title} {Flagellar
  hydrodynamics},\ }\href {https://doi.org/10.1137/1018040} {\bibfield
  {journal} {\bibinfo  {journal} {SIAM Rev.}\ }\textbf {\bibinfo {volume}
  {18}},\ \bibinfo {pages} {161} (\bibinfo {year} {1976})}\BibitemShut
  {NoStop}%
\bibitem [{\citenamefont {Bet}\ \emph {et~al.}(2017)\citenamefont {Bet},
  \citenamefont {Boosten}, \citenamefont {Dijkstra},\ and\ \citenamefont {van
  Roij}}]{bet_2017}%
  \BibitemOpen
  \bibfield  {author} {\bibinfo {author} {\bibfnamefont {B.}~\bibnamefont
  {Bet}}, \bibinfo {author} {\bibfnamefont {G.}~\bibnamefont {Boosten}},
  \bibinfo {author} {\bibfnamefont {M.}~\bibnamefont {Dijkstra}},\ and\
  \bibinfo {author} {\bibfnamefont {R.}~\bibnamefont {van Roij}},\ }\bibfield
  {title} {\bibinfo {title} {Efficient shapes for microswimming: From
  three-body swimmers to helical flagella},\ }\href
  {https://doi.org/10.1063/1.4976647} {\bibfield  {journal} {\bibinfo
  {journal} {J. Chem. Phys.}\ }\textbf {\bibinfo {volume} {146}},\ \bibinfo
  {pages} {084904} (\bibinfo {year} {2017})}\BibitemShut {NoStop}%
\end{thebibliography}%



\end{document}